\documentclass[aps,prd,twocolumn,amsmath,superscriptaddress,amssymb,showpacs,floatfix,nofootinbib,longbibliography]{revtex4-1}
\pdfoutput=1
\usepackage{graphicx}
\usepackage{bm}
\usepackage{times}
\usepackage{natbib}
\usepackage{slashed}
\usepackage{color}
\usepackage{aas_macros}
\usepackage{slashed}
\usepackage{lipsum}
\usepackage{subfigure}
\usepackage{multirow}
\usepackage{amsmath}
\usepackage{array} 
\usepackage{varwidth} 
\usepackage[table]{xcolor}

\newcommand{\be}{\begin{equation}}
\newcommand{\ee}{\end{equation}}
\newcommand{\bea}{\begin{eqnarray}}
\newcommand{\eea}{\end{eqnarray}}

\begin{document}


\title{First Observations of Solar Disk Gamma Rays over a Full Solar Cycle}

\author{Tim Linden}
\email{linden@fysik.su.se}
\thanks{\scriptsize \!\! 0000-0001-9888-0971}
\affiliation{Stockholm University and The Oskar Klein Centre for Cosmoparticle Physics,  AlbaNova, 10691 Stockholm, Sweden}
\affiliation{Center for Cosmology and AstroParticle Physics (CCAPP), Ohio State University, Columbus, Ohio 43210, USA}

\author{John F. Beacom}
\email{beacom.7@osu.edu}
\thanks{\scriptsize \!\! 0000-0002-0005-2631}
\affiliation{Center for Cosmology and AstroParticle Physics (CCAPP), Ohio State University, Columbus, Ohio 43210, USA}
\affiliation{Department of Physics, Ohio State University, Columbus, Ohio 43210, USA}
\affiliation{Department of Astronomy, Ohio State University, Columbus, Ohio 43210, USA}

\author{Annika H.~G. Peter}
\email{peter.33@osu.edu}
\thanks{\scriptsize \!\! 0000-0002-8040-6785}
\affiliation{Center for Cosmology and AstroParticle Physics (CCAPP), Ohio State University, Columbus, Ohio 43210, USA}
\affiliation{Department of Physics, Ohio State University, Columbus, Ohio 43210, USA}
\affiliation{Department of Astronomy, Ohio State University, Columbus, Ohio 43210, USA}

\author{\mbox{Benjamin J. Buckman}}
\email{buckman.12@buckeyemail.osu.edu}
\thanks{\scriptsize \!\! 0000-0002-0458-2402}
\affiliation{Center for Cosmology and AstroParticle Physics (CCAPP), Ohio State University, Columbus, Ohio 43210, USA}
\affiliation{Department of Physics, Ohio State University, Columbus, Ohio 43210, USA}

\author{Bei~Zhou}
\email{beizhou@jhu.edu}
\thanks{\scriptsize \!\! 0000-0003-1600-8835}
\affiliation{Center for Cosmology and AstroParticle Physics (CCAPP), Ohio State University, Columbus, Ohio 43210, USA}
\affiliation{Department of Physics, Ohio State University, Columbus, Ohio 43210, USA}
\affiliation{Department of Physics and Astronomy, Johns Hopkins University, Baltimore, Maryland 21218, USA}

\author{Guanying Zhu}
\email{zhu.1475@osu.edu}
\thanks{\scriptsize \!\! 0000-0003-0031-634X}
\affiliation{Center for Cosmology and AstroParticle Physics (CCAPP), Ohio State University, Columbus, Ohio 43210, USA}
\affiliation{Department of Physics, Ohio State University, Columbus, Ohio 43210, USA}


\begin{abstract}
The solar disk is among the brightest $\gamma$-ray sources in the sky.  It is also among the most mysterious. No existing model fully explains the luminosity, spectrum, time variability, and morphology of its emission. We perform the first analysis of solar-disk $\gamma$-rays over a full 11-year solar cycle, utilizing a powerful new method to differentiate solar signals from astrophysical backgrounds. We produce: (i) a robustly measured spectrum from 100~MeV to 100~GeV, reaching a precision of several percent in the 1--10~GeV range, (ii) new results on the anti-correlation between solar activity and $\gamma$-ray emission, (iii) strong constraints on short-timescale variability, ranging from hours to years, and (iv) new detections of the equatorial and polar morphologies of high-energy $\gamma$-rays. Intriguingly, we find no significant energy dependence in the time variability of solar-disk emission, indicating that strong magnetic-field effects close to the solar surface, rather than modulation throughout the heliosphere, must primarily control the flux and morphology of solar-disk emission. 
\end{abstract}

\maketitle


\section{Introduction}
\label{sec:introduction}

The Sun is a special astrophysical source. Its close proximity allows detailed studies critical to understanding other stars.  The ability to spatially resolve solar emission is especially important for probing high-energy, nonthermal processes, which can be highly local.  These processes reveal charged-particle acceleration and interactions in the Sun's complex, dynamic magnetic fields.  In addition, the ``space weather" induced by these processes affects Earth's atmosphere and our technological infrastructure, giving these studies practical as well as scientific importance~\cite{2011AdSpR..47.2059H}. 

The highest-energy processes are revealed by $\gamma$-ray observations up to $\sim$200 GeV, which correspond to charged parent particles at $\sim$10$\times$ higher energies.  Three processes produce GeV-range solar $\gamma$-rays.  Solar flares --- rapid, and well-localized, ejections of plasma from the solar surface --- can accelerate charged particles, producing $\gamma$-rays up to a few GeV~\cite{1987ApJS...63..721M, 1996A&AS..120C.299S, 2014ApJ...787...15A, 2014ApJ...789...20A, Pesce-Rollins:2015hpa, Omodei:2018uni}. These are easy to separate from other sources. Higher-energy $\gamma$-rays are produced through passive bombardment by cosmic rays.  Cosmic-ray electrons (and positrons) undergo inverse-Compton scattering (ICS) with solar photons, producing a $\gamma$-ray halo around the Sun~\cite{Moskalenko:2006ta, Orlando:2006zs, 2011ApJ...734..116A}.  Cosmic-ray protons (and nuclei) undergo hadronic interactions with matter in (and under) the solar photosphere, producing a bright disk~\cite{1965RvGSP...3..319D, 1966JGR....71.5778P, 2009arXiv0907.0557O, 2011ApJ...734..116A}.  For the halo, hadronic cosmic rays are irrelevant due to their small $\gamma$-ray production cross sections; for the disk, leptonic cosmic rays are irrelevant due to their small flux.  The available angular resolution lets us resolve the disk and halo components.  In this paper, we focus on the disk emission.

The Sun's $\gamma$-ray emission is dramatically affected by its magnetic fields.  Without magnetic fields, the disk emission would have two components.  At energies above $\sim$1~GeV, the $\gamma$-ray direction increasingly follows that of the parent cosmic ray.  Accordingly, only cosmic rays that graze the solar surface can interact and have the $\gamma$-rays escape~\cite{Zhou:2016ljf}.  The corresponding emission from the solar limb is too faint to be observed by the Fermi Gamma-Ray Space Telescope (Fermi).  Near 1 GeV, there is also a ``backsplash" component from the whole disk, as kinematics allow low-energy $\gamma$-rays to be emitted at a large angle relative to the parent cosmic ray~\cite{Zhu2020_TBS}.

Of course, the Sun does have magnetic fields.  Seckel, Stanev, and Gaisser (SSG~\cite{Seckel:1991ffa}) hypothesized that surface fields allow emission from the full disk even at high energies.  This requires the fields to deflect cosmic rays from incoming to outgoing before they interact and produce $\gamma$-rays.  This requires a ``Goldilocks" tuning between the solar magnetic field and gas density profiles. If cosmic rays are deflected too high in the solar atmosphere, they will not encounter enough matter to produce $\gamma$-rays.  If they are deflected too low, they will produce only $\gamma$-rays that are pointed into, and are subsequently absorbed by, the Sun.  To efficiently produce a broad $\gamma$-ray spectrum, SSG assume that cosmic rays impinging on the solar surface are funneled into magnetic flux tubes that mirror cosmic rays at the right depths.  The $\gamma$-ray flux predicted by SSG~\cite{Seckel:1991ffa} greatly exceeds that predicted for the solar limb~\cite{Zhou:2016ljf}. Detailed  simulations of cosmic-ray interactions within the solar atmosphere (including, e.g., solar composition, secondary production, coronal magnetic fields, etc.) have recently been completed by several groups~\cite{Mazziotta:2020uey, Li:2020gch}.

The steady-state emission from the solar disk and halo was not detected until 2008, in a re-analysis of EGRET data~\cite{2008A&A...480..847O}.  In 2011, an analysis with the Fermi Large-Aperture Telescope (Fermi-LAT) measured solar $\gamma$-ray emission between 100 MeV to 10 GeV, separating the disk and halo components~\cite{2011ApJ...734..116A}.  The disk itself was not resolved, but the halo was detected out to $\sim$20$^\circ$ from the Sun. Intriguingly, the solar disk flux exceeded the SSG prediction by a factor of $\sim$5.

Since 2011, new studies of Fermi-LAT data by our group have greatly improved these observations, identifying several new features.  Ng et al.~\cite{Ng:2015gya} showed that the solar disk $\gamma$-ray spectrum extends to $\sim$100~GeV and is significantly harder than the SSG predictions.  Additionally, this paper showed surprising evidence for time variability, finding that the $\gamma$-ray flux is strongly anti-correlated with solar activity.  Linden et al.~\cite{Linden:2018exo} took advantage of the high angular resolution of the Fermi-LAT above 10~GeV to perform the first resolved study of the disk, finding that the bright, hard-spectrum emission observed during solar minimum is produced predominantly in the Sun's equatorial plane, while the softer emission from polar regions is constant over the solar cycle.  Tang et al.~\cite{Tang:2018wqp} showed that the solar-minimum spectrum scales roughly as $E^{-2.2}$ from 100 MeV to beyond 100 GeV, except for a deep dip between 30--50~GeV, which is unexplained.

In this paper, we utilize a powerful new technique to separate solar signals from astrophysical backgrounds, which allows us to significantly increase statistics and minimize systematic uncertainties.  Using this model, we perform the first analysis of solar-disk $\gamma$-rays over a full solar cycle (2008--2020), including the first measurements below 1~GeV since 2011.  This large dataset allows us to measure the energy dependence of the anticorrelation between the disk flux and solar activity.  The long baseline also provides the statistics needed to search for short-period variations.  Because our observations happen to start at solar minimum, we are able to compare the emission between two different minima, over which the polarity of the Sun's dipole field has reversed.  This allows us to compare the morphology and spectrum of the emission across the solar disk during periods with opposite polarities. 

The outline of this paper is as follows. In Sec.~\ref{sec:methodology}, we describe our models of the solar disk, solar ICS halo, and astrophysical backgrounds. In Sec.~\ref{sec:results}, we measure the solar-disk flux, spectrum, time variability, and morphology. In Sec.~\ref{sec:conclusions} we interpret our results. In the appendices, we further examine our ICS halo and solar flare models. Overall, our results significantly extend the measurement of solar-disk $\gamma$-ray emission at low energies, complementing our work at TeV energies with the High-Altitude Water Cherenkov Observatory (HAWC) collaboration~\cite{Albert:2018vcq, Nisa:2019mpb}.

\begin{figure*}[t]
\includegraphics[width=1.97\columnwidth]{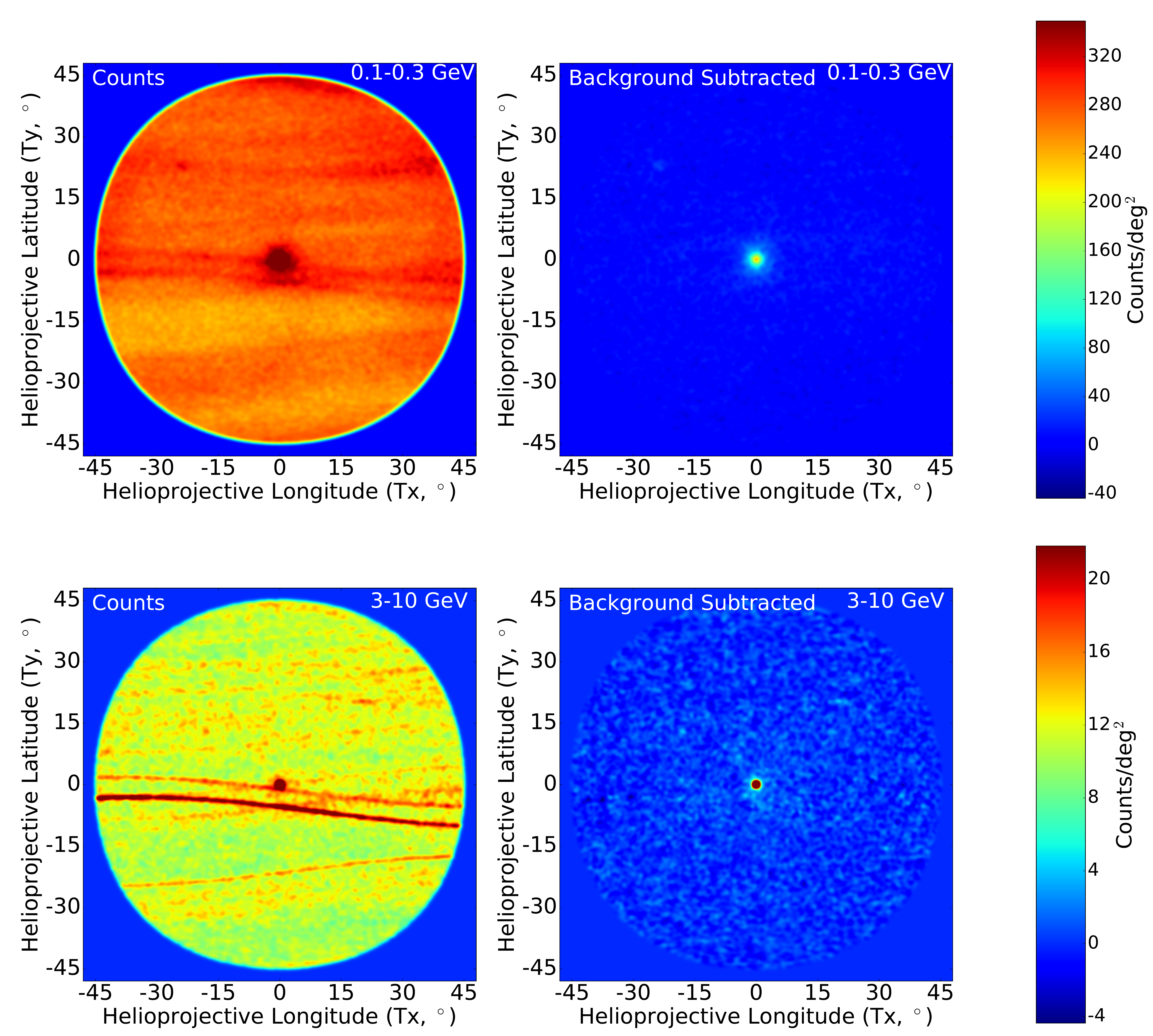}
\caption{{\bf Left:} $\gamma$-ray count maps in helioprojective coordinates over 11~years of Fermi-LAT data (Mollweide projection). Periods with significant Fermi-LAT flares are removed. Results are shown for 100--316~MeV (top) and 3.16--10~GeV (bottom). The outer ring is an edge effect due to pixels that are partially in the ROI. {\bf Right:} Residual $\gamma$-ray emission maps after the empirically determined background is subtracted. The color bars apply to both left and right panels. Our background-removal methodology efficiently removes both diffuse emission (the haze in the left panels) and point sources (the streaks in the left panels), leaving only the solar emission (disk and halo components) and statistical noise.}
\label{fig:bg}
\end{figure*}


\section{Data Analysis Methodology}
\label{sec:methodology}

We develop a novel, sophisticated methodology to measure the solar-disk $\gamma$-ray flux, improving on our previous analyses~\cite{Ng:2015gya, Tang:2018wqp, Linden:2018exo} in several ways. Most importantly, we use a data-driven model for the astrophysical $\gamma$-ray background that optimally isolates solar contributions and automatically accounts for systematic uncertainties in Fermi point-source determinations and diffuse background models.  We utilize and compare several methods to remove the contributions of solar flares to the disk flux.  Finally, we have developed and implemented detailed models for $\gamma$-ray production via the ICS solar halo, as well as the $\gamma$-ray production from the Moon. These technical improvements allow us to accurately determine the solar disk $\gamma$-ray flux down to 100~MeV. 


\subsection{Fermi-LAT Data Selection}
\label{subsec:dataselection}

We utilize 11.4~yr of P8\_V3 Fermi-LAT data, from August 2008 to February 2020 (MET 239557417--603960406), including SOURCE-class events from 100~MeV to 100~GeV with zenith angles $<$90$^\circ$. We use standard selection cuts. We divide the data into eight equally spaced logarithmic energy bins per decade. We calculate the exposure in 10 equal bins of the instrumental coordinate $\phi$, which describes the azimuthal angle surrounding the Fermi-LAT boresight ({\tt phibins=10}). This choice is not typically necessary in Fermi-LAT studies, but is important here as the orientation of Fermi's solar panels biases solar observations to regions near $\phi = 0$.

We analyze the data in helioprojective (Sun-centered, with the North pole defined by the Sun's spin axis) coordinates by separately transforming $\gamma$-ray events and the Fermi-LAT exposure. For events, we use the {\tt sunpy} code~\cite{2020ApJ...890...68S} to convert every $\gamma$-ray into solar coordinates, utilizing the exact time of each event. For the exposure, we use the Fermi-LAT tools {\tt gtltcube} and {\tt gtexpcube2} to calculate the exposure in J2000 coordinates over small (8640-second) time bins. In each bin, the Sun moves only $\sim$0.1$^\circ$ in J2000 coordinates. We convert this exposure map to helioprojective coordinates using the average time in each bin. This time binning is sufficient for our analysis, as the exposure is smooth on 0.1$^\circ$ scales.


\subsection{Fermi-LAT Solar Exposure Since March 2018}

Solar analyses are complicated by a solar-panel malfunction that affected Fermi on 16 March 2018. While this issue does not significantly impact the vast majority of Fermi science goals, the necessity of locking the fixed solar panel onto the Sun (to power the instrument) decreases the solar exposure, as the Fermi-LAT instrument is inclined relative to the solar panels.  For data beginning in March 2018, the monthly exposure is approximately half that recorded in previous periods. This is taken into account in our analyses of solar exposures and instrumental response functions. However, it worsens the statistical uncertainties in the most recent data.


\subsection{Removal of Solar Flares}
\label{subsec:timesets}

To isolate the effects from transient solar flares from the steady-state disk flux, we prepare three subsets of the data.  In the first, we apply no cuts, capturing the total $\gamma$-ray flux from the solar disk, including both active emission from solar flares and passive emission from hadronic cosmic-ray interactions. 

In the second, we apply a cut using the LAT Significant Flares List\footnote{https://hesperia.gsfc.nasa.gov/fermi/lat/qlook/lat\_events.txt}, removing events within 8640~s (2.4~hr) of a flare, the same size binning we use for exposure.  We  conservatively remove all events/exposure for which \emph{any} part of the bin overlaps with the full duration of the flare.  This cut removes only $\sim$0.36\% of our total solar exposure but significantly lowers the solar flux by removing several bright flares.

In the third, we additionally apply a cut based on keV--MeV data from the Fermi Gamma-ray Burst Monitor (GBM), using the GBM Solar Flare List~\footnote{https://hesperia.gsfc.nasa.gov/fermi/gbm/qlook/fermi\_gbm\_flare\_list.txt}.  To be conservative, we cut all $\gamma$-ray events that occur within \mbox{86400s (1 day)} of a flare from either GBM or LAT.  This removes a significant fraction (41.3\%) of the solar exposure and allows us to test whether sub-threshold flares impact the solar $\gamma$-ray spectrum.


\subsection{Astrophysical Background Modeling}
\label{subsec:backgroundmodel}

We produce a data-driven model of the astrophysical background produced by non-solar events. The approach is simple to describe but computationally demanding. The key concept is the following. While the motion of the Sun relative to distant astrophysical sources adds significant technical complexity, it also allows the astrophysical background flux to be directly measured during periods when the Sun is not present.

To characterize the astrophysical backgrounds, we begin with the helioprojective coordinates of each $\gamma$-ray recorded by Fermi-LAT as well as the finely binned (2.4-hr) exposure files. We remove all events and exposure within 45$^\circ$ of the current solar position.  To synchronize the event selection with our exposure calculation, we keep or remove Fermi-LAT events based on average right ascension and declination values of the Sun in each 2.4-hr bin.  Last, we convert the counts and exposure maps back to J2000 coordinates, producing a flux map of the entire sky during periods when the Sun was far away. 

This methodology significantly outperforms previous approaches.  Because our model is data driven, it is insensitive to uncertainties in diffuse background modeling and point-source estimation. Additionally, our model automatically includes effects stemming from the Fermi-LAT angular and energy resolution. Compared to previous data-driven analyses (for example, the Fake-Sun methods employed in~\cite{Ng:2015gya, Tang:2018wqp, Linden:2018exo}), this technique produces a significantly larger statistical sample, in effect using a continuum of Fake Suns instead of a handful.  Even more, we produce a Sun-centered skymap that contains nearly every Fermi-LAT $\gamma$-ray. Our eventual analysis includes 49600 photons that are assigned to the solar disk template, including 8400 above 1~GeV and 513 above 10~GeV, providing a significant statistical handle on the solar disk flux, time variability, and morphology.

Figure~\ref{fig:bg} shows the results of this method in two example energy bins, 100--316 MeV and 3.16--10 GeV.  The $\gamma$-ray flux in helioprojective coordinates is shown before and after background subtraction. Bright diffuse emission (with a flux nearly half that of the solar disk at low energies) as well as bright point sources (observed as tracks that streak past the Sun) are entirely removed, leaving only the solar (disk and halo) components, plus statistical noise. We note that each track corresponds to a single bright $\gamma$-ray source, with the direction of the track depending on the sources exact motion through helioprojective coordinates.


\subsection{Additional Model Considerations}
\label{subsec:moon_and_ics}

There are three additional complications that affect our background methodology (and those of previous papers).

First, while our method produces a robust model for all sources that do \emph{not} move in the J2000 coordinate system, it produces a skewed model of any source that moves with respect to background stars. The two important sources in this class are (1) the Moon and (2) the diffuse solar ICS halo. While the Moon moves compared to the Sun, the solar ICS halo is stationary in helioprojective coordinates. Thus, the two components are treated using different techniques. 

To account for the Moon's emission, we slightly amend the above analysis to additionally remove any $\gamma$-rays that are observed within 20$^\circ$ of the Moon in each 2.4-hr time bin.  While the Sun moves only 0.1$^\circ$ during this time period, the Moon moves approximately 1.2$^\circ$. However, this is still small compared to both the angular dependence of the Fermi-LAT exposure as well as the size of the point-spread function (PSF) at the $\sim$100~MeV energies where the Moon is bright. Our amended model thus produces an analysis of the astrophysical background in regions where neither the Moon nor Sun are present. We model the Moon's $\gamma$-ray emission by utilizing {\tt gtmodel} to produce a source with radius 0.26$^\circ$ at the current position of the Moon in each 8640-s exposure bin. We fit the flux of the Moon component in our likelihood analysis. This enables us to avoid throwing out approximately 25\% of our exposure, which is taken during periods when the Moon lies within 45$^\circ$ of the Sun.

Solar ICS emission is concentrated within 45$^\circ$ of the Sun, a region that is already removed from our analysis of the astrophysical backgrounds. We build a solar ICS emission model utilizing the {\tt StellarICs} code~\cite{Orlando:2013nga, Orlando:2013pza}, which we slightly modify to include updated local interstellar spectra for cosmic-ray electrons and positrons~\cite{Bisschoff:2019lne}. The modulated spectra are calculated using the standard force-field approximation~\cite{Gleeson:1968zza}, which assumes separability of the heliospheric diffusion coefficient into radial- and energy-dependent functions. For the radially-dependent force-field potential, we use Eq.~(7) of Ref.~\cite{Moskalenko:2006ta}, which is derived from the radial dependence of the cosmic-ray mean free path for solar cycles 20/22~\cite{fujii2005spatial}.  The largest uncertainty in our study, particularly at low energies, comes from our modeling of the ICS emission.  We provide more details in Sec.~\ref{subsec:icsmodels}.

Because the astrophysical background template is based on real data, it already includes the effects of the PSF. However, the solar disk, solar ICS halo, and Moon templates must be smeared by the Fermi-LAT angular resolution. The Moon's $\gamma$-ray flux is smeared using {\tt gtmodel}. For the Disk and ICS flux, we note that the PSF near the Sun may systematically differ from the average Fermi-LAT PSF due, e.g., to the extreme $\phi$-dependence of solar observations. Thus, we calculate the PSF in each 2.4~hr exposure using {\tt gtpsf}. The solar PSF is then weighted by the solar exposure in each bin. Because the exact structure of the PSF farther from the Sun is less critical for our study, we utilize a PSF calculated at the solar position throughout the ROI. The average energy resolution of Fermi-LAT events ($\sim$10\%) is smaller than the energy binning of our analysis and the solar spectrum is relatively smooth, so we can safely neglect simulating the energy dispersion.

The second complication stems from variable $\gamma$-ray sources, which may have different luminosities when they are close to the Sun, producing residuals in the $\gamma$-ray analysis. However, this does not significantly affect our results for two reasons. First, the 11-yr lever arm of our analysis means that each source has made multiple passes near the Sun, decreasing the effects of single source flares by a factor of $\sim$10. Second, these source residuals produce bright streaks across the solar position, which are not degenerate with the spherically symmetric solar emission, as seen in Fig.~\ref{fig:bg}. The combination of these factors implies that variable $\gamma$-ray sources have a very small effect on our determination of the solar disk flux.

The third complication is that the unique nature of the Sun makes it difficult to entirely eliminate systematic effects specific to the solar region. For example, because Fermi's solar panels are locked onto the Sun, the instrumental phase space of solar events is unique. This potentially affects the effective area, angular resolution, and energy resolution of solar events. Moreover, radiation or cosmic-ray effects on the instrument itself cannot be entirely ruled out. Notably, our previous study~\cite{Linden:2018exo} uncovered a systematic error in events near $\phi =0$, which has since been corrected in the latest Fermi-LAT data release.  Our thorough studies of systematics in Ref.~\cite{Linden:2018exo} suggest that any remaining issues must be very subtle.

\begin{figure*}[t]
\includegraphics[width=1.98\columnwidth]{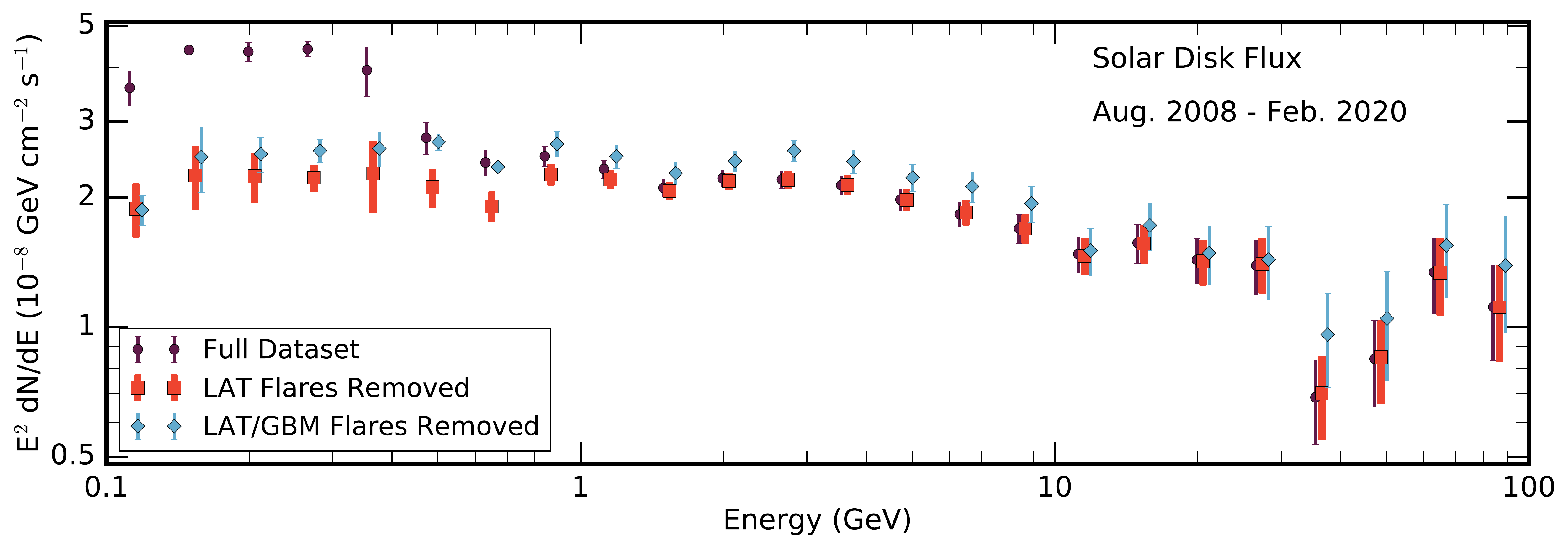}
\caption{Average solar-disk $\gamma$-ray spectrum over the full analysis period (Aug. 2008--Feb 2020). Results are shown for three choices of solar-flare cuts: no cuts (purple), removing high-significance Fermi-LAT detected flares (bold orange), and additionally removing high-significance Fermi-GBM flares (blue). Small offsets on the x-axis are made to improve visibility.  Removing Fermi-LAT flares is important below $\sim$1 GeV.  Removing Fermi-GBM flares is not important.  The GBM-cut fluxes are slightly higher because the steady-state solar-disk flux is dimmer during solar maximum, when more flares are present.  See text for details.}
\label{fig:spectrum}
\end{figure*}

To conclude, we are confident that our methodology goes far beyond previous analyses. This allows us to utilize nearly every solar $\gamma$-ray, regardless of the solar position compared to background $\gamma$-ray sources, while simultaneously removing all background contributions with high fidelity. 


\subsection{Fitting the ICS Component}
\label{subsec:icsmodels}

Modeling the background from the solar ICS halo is especially challenging, both because it is stationary in heliospheric coordinates and because its emission extends very close to the disk. The degeneracy between disk and halo emission is particularly bad below 1~GeV, where the Fermi-LAT angular resolution worsens significantly.

The uncertainty is magnified by our ignorance of the ICS morphology, which prevents us from reliably extrapolating the ICS flux from positions far from the Sun, where the disk flux is irrelevant. Instead, we test multiple models of the ICS morphology and marginalize our results over them. We begin with three ``seed" models for the ICS morphology that use modulation potentials defined at 1 AU of \mbox{$\Phi_0 = \{0, 400, 1000\}~{\rm MV}$}. These choices liberally span the range of modulation potentials that are compatible with local cosmic-ray observations. We note that we do not intend to measure the modulation potential of the Sun in this analysis --- instead we develop a model where the measured solar $\gamma$-ray flux is independent of the ICS modulation model. In a forthcoming publication~\cite{Linden2021_TBS}, we will measure and analyze the ICS component in detail.

In our default analysis, we divide the ICS templates for each modulation model into seven angular components, spanning \{0--1, 1--2, 2--4, 4--6, 6--8, 8--10, and 10--45\}$^\circ$ from the solar position. These angular regions are defined \emph{before} the emission is smoothed by the PSF. Thus, these templates represent physical emission on each angular scale. Utilizing this binning, our three distinct ``seed" models provide different ICS angular profiles \emph{within} each radial bin, but the floating normalizations of each independent bin allow each model to have similar global properties on angular scales much larger than a bin. Our goal, in this analysis, is to show that all three seed models produce the same disk flux, which will indicate that small scale features in the ICS morphology do not affect our results. We note that these templates are allowed to float to different best-fit values within each energy bin. 

This approach has two benefits. First, it accounts for uncertainties in the ICS halo profile near the Sun due to uncertainties in solar modulation. Second, and more subtly, it disentangles uncertainties in the relative normalization of the ICS and astrophysical background far from the Sun with the ICS and solar disk components near the Sun. While the ICS surface brightness is highest within $\sim$2$^\circ$ of the Sun, the large extent of the halo means that the average ICS $\gamma$-ray is produced $\sim$10--15$^\circ$ from the Sun. Thus, in a model with only one ICS component, the ICS normalization would be set by the low-surface brightness emission observed far from the Sun, which is then extrapolated to calculate the ICS flux at the solar position. The effect of this extrapolation is shown in Appendix~\ref{app:ics}.

We utilize several methods to control variation in the normalization of each ICS angular bin. These choices moderately affect the measured solar emission, but more significantly affect its uncertainty. In our default analysis, we conservatively allow the normalization of each ICS angular bin to float freely, demanding only that it be non-negative.  This produces robust error bars that encompass the systematic uncertainties of more fine-tuned normalization models. However, this potentially produces unphysical ``saw-tooth" features in the ICS angular profile. In Appendix~\ref{app:ics}, we show the results from physically motivated models that force the ICS normalization coefficients to either (1) be monotonic in angle, or (2) incur a renormalization penalty for adjustments that stray from unity. 

We stress two important results. First, the modeling uncertainties affect the solar disk spectrum only below $\sim$1~GeV. At higher energies, we robustly separate solar disk $\gamma$-ray emission from ICS and astrophysical emission regardless of our ICS modeling choices. Second, our choice to set extremely conservative constraints on the variation of the ICS modeling template implies that the uncertainty in our solar disk fits encompass both statistical errors as well as systematic errors from ICS halo mismodeling.

\subsection{The Solar Disk Component}
\label{subsec:disk}
Finally, we include a template for the solar disk. In our analysis, we assume a disk with a constant surface brightness out to a radius of R$_\odot$, which corresponds to $\sim$0.254$^\circ$. We note that this is smaller than the Fermi-LAT angular resolution at all but the highest $\gamma$-ray energies. In Section~\ref{subsec:morphology}, we investigate the solar disk $\gamma$-ray morphology in detail at energies above 10~GeV, where the solar disk emission can be resolved.


\subsection{Likelihood Analysis}
\label{subsec:likelihoodanalysis}

To measure the solar disk $\gamma$-ray flux, we employ a template-based likelihood minimization analysis using the public code {\tt iminuit}. We utilize an ROI of radius 45$^\circ$ centered on the Sun, binned with a HEALPix grid of nside = 512 (pixel size $\sim$0.11$^\circ$). We carry out our analysis in 24 independent bins from 100~MeV to 100~GeV, equally spaced in  logarithmic energy, so that we can measure the spectrum shape.

We fit the data with the four-component model described above: (1) a constant surface-brightness solar disk of radius 0.26$^\circ$, the size of the physical disk, (2) the multi-component ICS template, following Sec.~\ref{subsec:icsmodels}, (3) the Moon template from Sec.~\ref{subsec:moon_and_ics}, and (4) the background model described in Sec.~\ref{subsec:backgroundmodel}. While we allow the normalization of the astrophysical background to float, our method of constructing this template should force the normalization to unity.  Indeed, we find variations of $<$1\% at most energies. We also find that our analysis is robust to changes in the Moon and astrophysical background models, but can be affected by ICS modeling choices below 1~GeV (see Appendix~\ref{app:ics}).


\section{Measurements of the Solar-Disk Emission}
\label{sec:results}

Utilizing the methodology outlined above, we measure the solar disk flux and spectrum~(Sec.~\ref{subsec:flux}), time variability (Sec.~\ref{subsec:variability}), and morphology (Sec.~\ref{subsec:morphology}). 


\subsection{Flux and Spectrum}
\label{subsec:flux}

Figure~\ref{fig:spectrum} shows the $\gamma$-ray spectrum for three different choices of solar-flare cuts. In each case, we utilize our multi-binned ICS halo model, using a modulation potential of \mbox{$\Phi_0=400$~MV} at 1 AU. We show in Appendix~\ref{app:ics} that this choice does not affect our results.  We note several results common to each choice of solar-flare cut: (1) we produce a robust measurement of the solar disk $\gamma$-ray spectrum down to 100~MeV, the first since 2011~\cite{2011ApJ...734..116A}, with uncertainties of 5--10\% in 0.1--1 GeV, (2) we obtain excellent precision for the flux, reaching uncertainties of 2--5\% in the 1--10 GeV range, and (3) we reproduce the unexpected ``spectral dip" feature between 30--50~GeV, first identified in Ref.~\cite{Tang:2018wqp}. We note that these flux uncertainties do not include a $\sim$10\% uncertainty in the Fermi-LAT effective area, which affects all flux measurements produced from Fermi analyses~\cite{2009ApJ...697.1071A}. 

Importantly, we stress that these results include significant freedom in the global profile of the ICS emission component (which we find is the largest systematic uncertainty in the model) -- thus the error bars shown here are conservative and include both statistical and systematic contributions. Below $\sim$1~GeV, we find that the systematic uncertainties from the global ICS morphology are dominant (see Appendix~\ref{app:ics}).

Notably, our analysis finds that the solar spectrum is remarkably flat, following a power-law very close to E$^{-2.0}$ from 100~MeV to nearly 30~GeV. This is noteworthy and theoretically unexpected, because the parent cosmic-ray proton population is observed to have an E$^{-2.7}$ spectrum across the relevant energy range. Because the outgoing $\gamma$-ray spectrum from isotropic cosmic-ray interactions is equivalent to parent proton spectrum, our results imply that the probability of a proton interacting with the solar atmosphere and producing an outgoing $\gamma$-ray increases strongly ($\sim$E$^{+0.7}$) with energy. The mechanism that would produce such an feature is unknown.

Below 1~GeV, the results strongly depend on the solar-flare cuts.  In the ``Full Dataset," which includes all solar flares, we observe a pronounced spike in the flux between 100~MeV and 1~GeV.  Nearly 30\% of the total solar emission below 1~GeV is produced by a handful of intense flares (see Sec.~\ref{subsec:timesets}). More importantly, however, we obtain the same $\gamma$-ray spectrum regardless of whether we choose to remove only the brightest ``LAT-detected" flares, or to more conservatively remove all ``LAT- or GBM-detected" flares. This implies that sub-threshold flares contribute negligibly to the total $\gamma$-ray flux.  Thus, removing the LAT-detected flares (encompassing only $\sim$0.3\% of the total exposure) sufficiently isolates the steady-state solar disk emission, and we use that in what follows.

\begin{figure}[t]
\includegraphics[width=0.98\columnwidth]{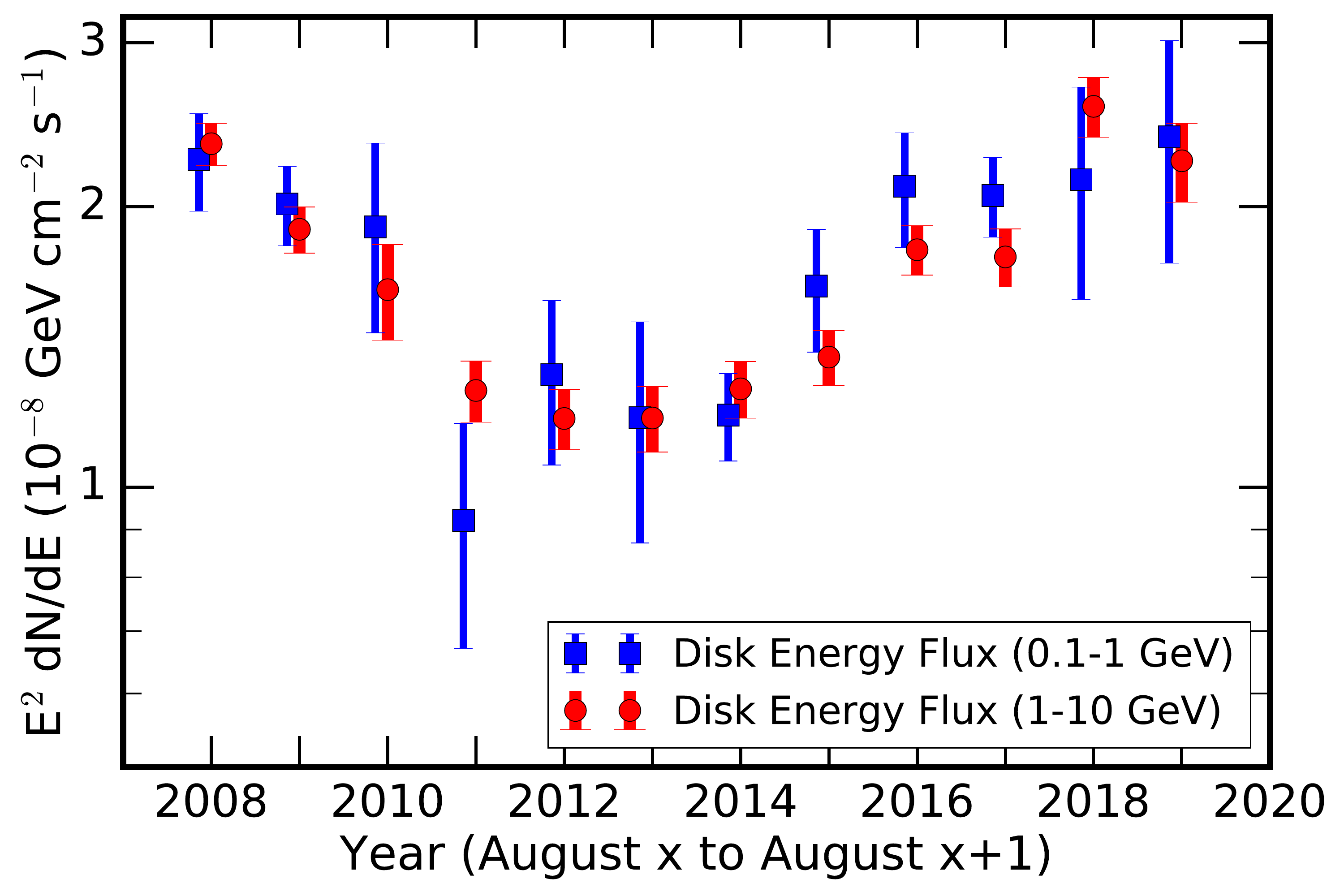}
\caption{Solar-disk $\gamma$-ray flux for 1--10~GeV (red) and 0.1--1~GeV (blue) in year-long analysis windows beginning approximately on August 4 of each year (the last bin ends in Feburary 2020).  The points are slightly offset for clarity.  We find strong evidence that the 11-yr variability, previously shown to anticorrelate with sunspot number~\cite{Ng:2015gya}, is energy-independent, which is surprising.}
\label{fig:time_variability}
\end{figure}

Focusing on the emission above 1~GeV, we note that the solar disk flux is larger during when ``LAT- or GBM-detected" flares are removed compared to when only ``LAT-detected" flares are removed.  This is an artifact of our selection criteria. The ``LAT- or GBM-detected" flare cut removes 41.3\% of the solar exposure, predominantly during solar maximum, when the solar flare rate is highest and the steady-state disk emission is faintest~\cite{Ng:2015gya, Tang:2018wqp, Linden:2018exo}. Thus, the ``GBM-detected" cut biases the 11-yr disk flux towards periods when the disk is brighter. In App.~\ref{app:flares}, we analyze the $\gamma$-ray flux above 1~GeV in yearly bins, finding that removing GBM-detected solar flares does not affect the solar $\gamma$-ray flux during any particular phase of the solar cycle, but instead re-weights the emission from different time periods.  We cannot, however, rule out the interesting possibility that an underlying variable affects both the solar flare cycle and the steady state flux of the solar disk.


\subsection{Time Variability}
\label{subsec:variability}

Our detailed background modeling and choice to remove only significant LAT-detected flares allows us to retain nearly 99.7\% of solar-disk exposure. Using this vast dataset, we can probe both long- and short-timescale variability.


\subsubsection{Long-Term Variability over a Solar Cycle}

Figure~\ref{fig:time_variability} shows the $\gamma$-ray flux in year-long bins. In each bin, we have sufficient data to refit our four-component model (astrophysical background, Moon, solar ICS halo, and solar disk), allowing us to account for variations in background sources or ICS emission. While our models fit emission over the full energy range, here we focus on two different energy scales. The first covers the range 1--10~GeV, where our measurements are the most precise; the second covers the range 0.1--1~GeV, where the effects of time-dependent solar modulation are expected to be greatest. At higher energies the time-variation appears consistent with the lower-energy data, but the statistical uncertainties become significantly worse, we provide a more detailed analysis in Appendix~\ref{app:highenergyvariability}. 

\begin{figure}[t]
\includegraphics[width=0.98\columnwidth]{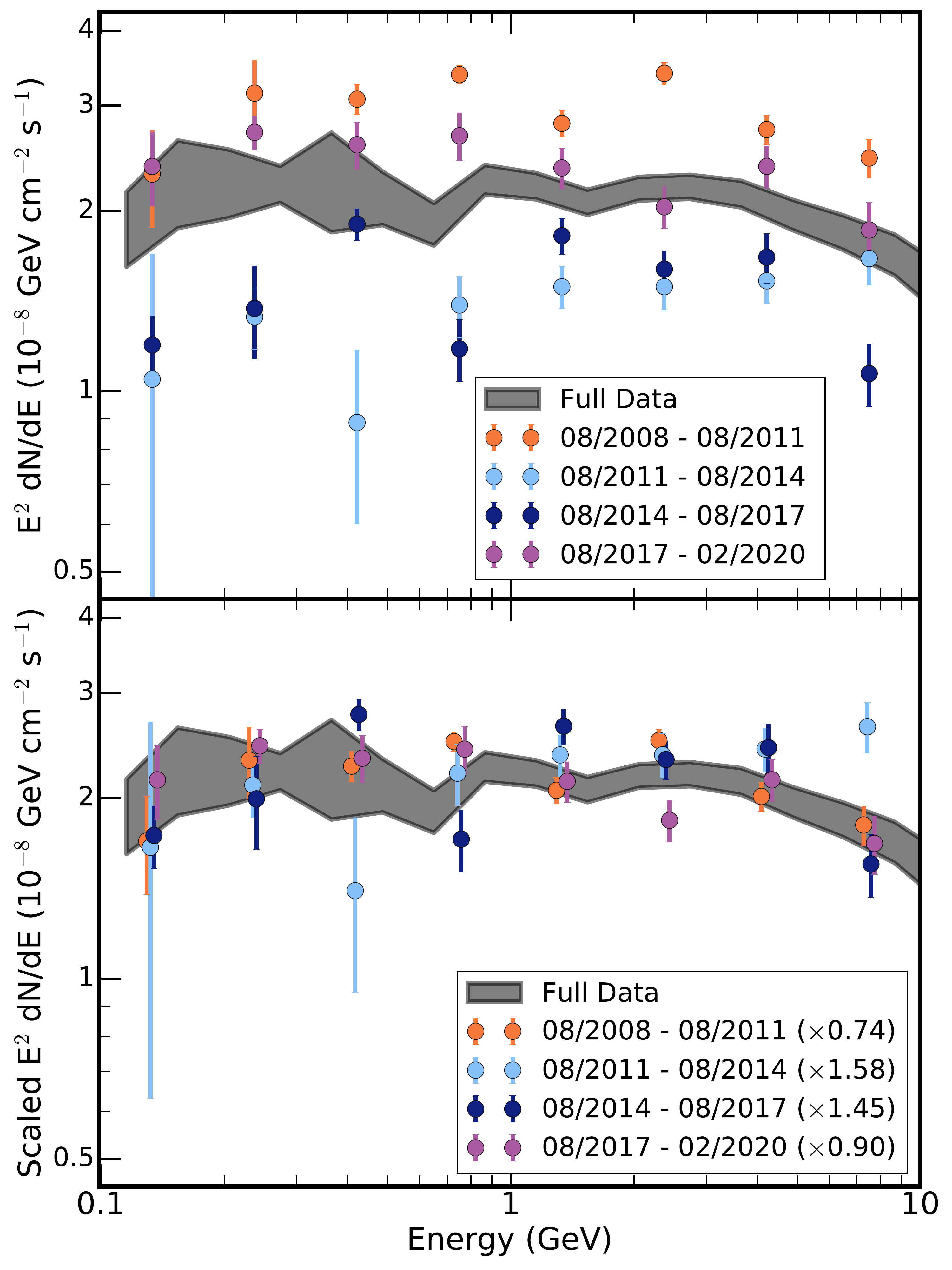}
\caption{{\bf Top:} Low-energy solar-disk $\gamma$-ray spectrum in four periods, each of duration approximately three years, as marked, measured utilizing our standard analysis routine and compared to the average flux over the full observation period. For clarity, we display results using only four bins per decade. {\bf Bottom:} Same as above, but the $\gamma$-ray flux has been renormalized in each time period to match the average flux in the four bins closest to 1~GeV. Our results do not indicate strong spectral differences between the solar minimum periods (represented by the first and last time periods), and the solar maximum time period (represented by the middle two temporal periods).}
\label{fig:spectrum_variability}
\end{figure}

We find a highly significant anti-correlation between the solar-disk flux and solar activity, echoing previous results~\cite{Ng:2015gya, Tang:2018wqp}. Specifically, we find a bright peak in the $\gamma$-ray emission in 2008 (near solar minimum), which falls to a valley during solar maximum (and with no appreciable changes during the 2013 heliospheric polarity flip), before rising again in 2020 (near solar minimum). Importantly, we find that the solar-disk fluxes during the prior and present solar minima are similar, despite their opposite polarities for the solar magnetic field (see Appendix~\ref{app:highenergyvariability} for a dedicated discussion at high energies). The larger error bars after March 2018 correspond to a decrease in Fermi-LAT exposure.

Figure~\ref{fig:spectrum_variability} shows the low-energy solar-disk $\gamma$-ray spectrum in four different temporal periods. The first roughly corresponds to solar minimum with polarity $A < 0$, the second to solar maximum with $A < 0$, the third to solar maximum with $A > 0$ and the final to solar minimum with $A > 0$. While we have analyzed the data in each temporal period utilizing our standard energy binning of eight logarithmic bins per decade, for visual clarity we show the resulting spectra after using a weighted average to decrease this to four bins per decade.

\begin{figure*}[t]
\includegraphics[width=1.95\columnwidth]{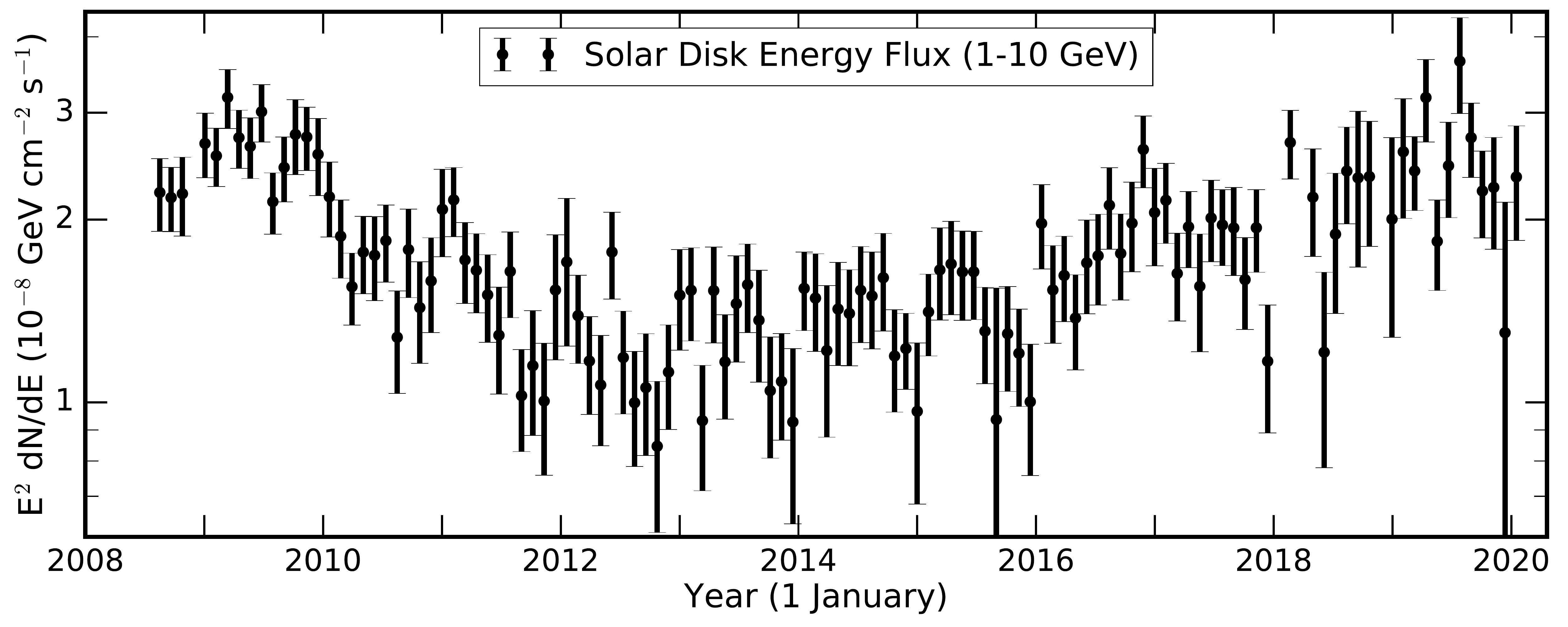}
\caption{Total solar-disk $\gamma$-ray flux between 1--10~GeV (where the flux is best measured) in $\sim$35-day (3~Ms) analysis windows. The 11-year periodicity of the solar-disk signal is clearly visible, but no significant periodicity is visually obvious on smaller timescales.}
\label{fig:time_variability_month}
\end{figure*}

Intriguingly, our analysis does not find any significant spectral variations once the overall change in the flux normalization is accounted for. Such a feature might have been expected, as the degree of cosmic-ray solar modulation near the Earth is a strongly energy-dependent phenomenon~\cite{Cholis:2020tpi}. The lack of such a feature implies that magnetic fields in the solar atmosphere, rather than the solar wind or cosmic-ray modulation far from the Sun, dominate the time-variability of the solar-disk emission.


\subsubsection{Monthly Variability}

Figure~\ref{fig:time_variability_month} shows the time variability of the 1--10~GeV flux over even smaller time ranges, of approximately 3$\times$10$^{6}$ seconds (34.7 days) per bin. In these small time bins, the quantity of data is insufficient to precisely remove the degeneracy between solar disk emission and the ICS emission in the inner (0--1$^\circ$) angular bin. Thus, in this sub-subsection, we fix the morphology of the ICS emission template (the relative flux in each ICS bin) to its best-fit value from our 11.4-yr analysis. We do, however, allow the normalization of this ICS component, along with the normalizations of the solar disk, Moon, and astrophysical background, to float independently in each energy and temporal bin. In our main analysis, we find that the ICS morphology is only slightly degenerate with the best-fit disk flux, indicating that this choice should only marginally affect the results here. 

We typically obtain $>$10$\sigma$ detections of solar $\gamma$-ray emission in each time bin.  We recover the 11-yr cycle found in Fig.~\ref{fig:time_variability}. However, there are no visually obvious variations on the shorter timescales probed here.  This implies that the short duration magnetic outbursts (e.g., solar flares, coronal mass ejections) are unlikely to determine the solar disk $\gamma$-ray flux. 


\subsubsection{High-Frequency Variability}

Finally, we utilize the precise timing information of each solar $\gamma$-ray event to search for high-frequency periodic variations in the flux. Such signals might provide important evidence into the regions of the Sun that produce bright solar emission. Additionally, this analysis provides an additional handle to test the contribution of solar flares to the steady-state solar disk $\gamma$-ray data.

One question of particular interest is whether certain regions of the Sun are more important than others for $\gamma$-ray production. Due to the solar rotation, this may manifest as a periodicity with a period of approximately 27 days (1 Bartels rotation, which characterizes the Sun's apparent rotation as seen from the moving Earth). The large Fermi-LAT dataset would, in principle, make it possible to fold the Fermi-LAT counts and exposure into the solar-phase space. However, because the solar atmosphere is not solid (polar regions rotate more slowly than equatorial regions), it is difficult to fold multiple solar rotations without smearing out any transient signal.

\begin{figure*}[t]
\includegraphics[width=1.98\columnwidth]{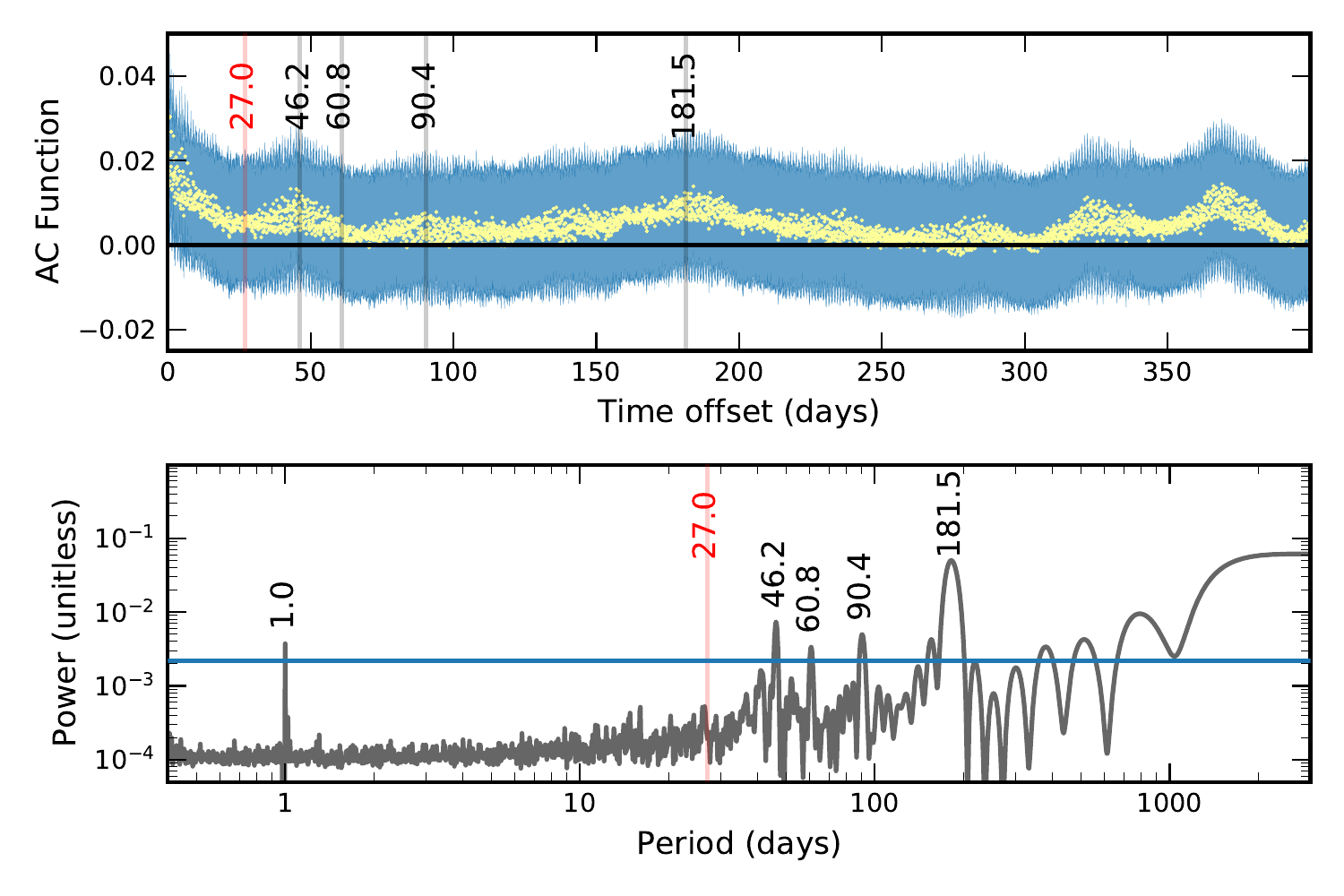}
\caption{Tests of time variability of the solar-disk $\gamma$-ray flux on a wide ranges of scales.  {\bf Top:} Autocorrelation function for offsets less than 400 days. The vertical lines denote the potential periodicities shown in the bottom panel. {\bf Bottom:} Power spectrum for offsets less than 3000 days. Peaks above the 99\% false alarm level (blue line) are denoted by the period of the peak value. The red vertical line at 27 days corresponds to 1 Bartels rotation, and indicates a region where we may expect a periodic signal due to solar rotation.} 
\label{fig:autocorrelation}
\end{figure*}

To maximize the time resolution of our analysis, in this sub-subsection we employ a different analysis technique. We begin with the $\gamma$-ray data above 1~GeV, binned into 8640-s periods, which corresponds to our choice of exposure binning. We select all events observed within 2$^\circ$ of the solar position, noting that the average 68\% confidence interval of the PSF for $\gamma$-rays above 1~GeV is always smaller than 1$^\circ$. This data-sample then contains the sum of all emission within 2$^\circ$ of the Sun coming from the solar disk, solar ICS halo, Moon, and astrophysical background. 

To isolate the variability of the solar disk, we note the following: (1) we can safely ignore the flux contribution from the Moon, because the 1~GeV minimum energy cutoff removes the vast majority its $\gamma$-ray emission, (2) we assume that the ICS emission is relatively time-invariant, moreover, above 1~GeV, the emission from the solar disk is significantly larger than the ICS flux within $\sim$2$^\circ$ of the solar position~\cite{2011ApJ...734..116A}

We note, however, that the astrophysical background can be extremely variable in helioprojective coordinates, as the Sun regularly (and periodically) moves past regions of high and low diffuse and point-source emission. To account for this, in each 8640-s exposure bin, we calculate the number of $\gamma$-ray counts expected from the astrophysical background.  We set this as the expectation value for the background flux during that temporal bin. Utilizing this energy cut and ROI, we observe 20661 counts compared to a background expectation of 9074 counts, implying that our signal is roughly 60\% produced by the solar disk. Because our model of the data is very accurate (see, e.g., Fig.~\ref{fig:bg}) we expect the difference between the background data and the background model to be mostly time-independent. The solar emission, on the other hand, may have significant variability on a variety of timescales. 

To measure the variability of the solar-disk flux, and compute its autocorrelation function, we first calculate the median flux along with non-Gaussian uncertainties ($\pm 1\sigma$ values) using a Bayesian technique. We assume that the solar and background fluxes are each Poisson-distributed and that the priors on each components are uniform in the logarithm.

To calculate the autocorrelation function at a time offset $t$, we calculate the Spearman correlation coefficient for the collection of all flux pairs separated by time $t$. We use the Spearman correlation coefficient because it is robust to outliers. We account for flux uncertainties through bootstrapping. Specifically, we calculate the autocorrelation function 10000 times, each time using a different light curve consistent with the non-Gaussian flux uncertainties. We then extract the mean and standard deviation of the autocorrelation function at each time offset. We calculate the autocorrelation function out to 3000-day offsets, slightly below the length of our total observation period.  We then calculate the power spectrum of the solar disk light curve by calculating the Lomb-Scargle periodogram (LSP) of the autocorrelation function using the public code {\tt astropy}~\cite{2013A&A...558A..33A}. The LSP takes into account the uncertainties of the autocorrelation function.  


Figure~\ref{fig:autocorrelation} shows the results of this analysis, including both the autocorrelation function and its power spectrum.  This analysis confirms significant power in the autocorrelation function on the longest timescales, finding significant and constant power at all periods exceeding 1000~days. This confirms the significant long-term periodicity (consistent with a solar cycle) is found in our primary dataset, utilizing a separate analysis technique without ICS or Moon modeling.

We find several additional periodicities that are statistically significant. However, most of these are explained  due to a combination of astrophysical and instrumental factors. We find a significant peak at periods $\sim$181.5~days, which is consistent with the two passages of the Sun across the galactic plane per year.  The analysis of this sub-subsection does a poorer job subtracting the astrophysical background.   The increased variability in our background signal as the Sun passes the bright galactic plane is likely to induce a periodic signal. We also find significant periodicities at 60.8 and 90.4 days that are well explained as higher harmonics of the physical 181.5~day periodicity.

Additionally, we find a significant line at 46.2 days, which is connected with the 53-day orbital procession period of Fermi-LAT~\footnote{https://fermi.gsfc.nasa.gov/ssc/observations/types/exposure/}. This 53-day orbital precision period occurs with respect to J2000 coordinates, and the Sun's motion across 14\% of the sky over that period decreases the orbital precession period (with respect to the Sun) to $\sim$45 days.

Finally, we find reasonably significant evidence for a 1~day periodicity of unknown origin. We note several key features of this periodicity: (1) it is approximately a 4$\sigma$ effect, (2) our flux bins are 0.1~days, implying that binning effects may be present, though our analysis does not necessarily fail for periodicities near our temporal binning, (3) the periodicity is exactly 1~day (within the few minute resolution of our scan), (4) the periodicity is found in the instrumental exposure (at the level of a $\sim$1\% fluctuation), and does not appear to be present in the counts data, (5) a one-day periodicity has been found in other Fermi-LAT observations, though these studies have not definitively determined its origin\footnote{https://fermi.gsfc.nasa.gov/ssc/data/analysis/LAT\_caveats\_temporal.html}. Combining these factors, we believe this effect to be of unknown instrumental origin. 

Most importantly, we find no evidence of significant power in the variability spectrum on periods between 24--34 days, which correspond to the rotation periods of the solar atmosphere around the equatorial and polar regions of the Sun, respectively. This strongly constrains models where certain ``hotspot" regions of the solar atmosphere could persist and produce significant $\gamma$-ray fluxes over multiple solar rotations.


\subsection{Morphology}
\label{subsec:morphology}

Finally, we utilize the high Fermi-LAT angular resolution above 10~GeV to study the $\gamma$-ray emission morphology across the solar disk. This analysis mostly repeats Ref.~\cite{Linden:2018exo}, but includes two more years of data from the recent solar minimum. Compared to the rest of the paper, this subsection includes six extra months of data (up until 29 August 2020), because here it is easy to do and because maximizing the total exposure is important due to the scarcity of high-energy $\gamma$-rays.

Unlike other analyses in this paper, in this section we do not perform any of the background subtraction methods described in Section~\ref{sec:methodology}. Instead, we copy the techniques of Ref.~\cite{Linden:2018exo}, to which we refer the reader for details. The reasons are two-fold. First, every $\gamma$-ray above 10~GeV is very well localized to within $\sim$0.3$^\circ$ of its true location. The solar disk intensity in this ROI is more than 10$\times$ brighter (at the highest energies, 100$\times$ brighter) than both ICS and astrophysical backgrounds, rendering their modeling irrelevant. Second, with such a high angular resolution, the Sun itself further inhibits the background. $\gamma$-rays from astrophysical objects will be blocked by the Sun, producing a hole in astrophysical $\gamma$-ray emission that fills roughly 25\% of the 0.5$^\circ$ ROI we employ. 

In our previous analysis, we found statistically significant, and completely unexpected, evidence for two morphological components of solar-disk $\gamma$-ray emission. The first emanates predominantly from the Sun's polar regions, and has a flux that is nearly independent of the solar cycle. The second emanates predominantly from the Sun's equatorial plane, and has a flux that peaks strongly during solar minimum and disappears during the remainder of the solar cycle.  The equatorial component produces a harder $\gamma$-ray spectrum that continues to $\gamma$-ray energies above 100~GeV. 

\begin{figure*}[t]
\includegraphics[width=1.75\columnwidth]{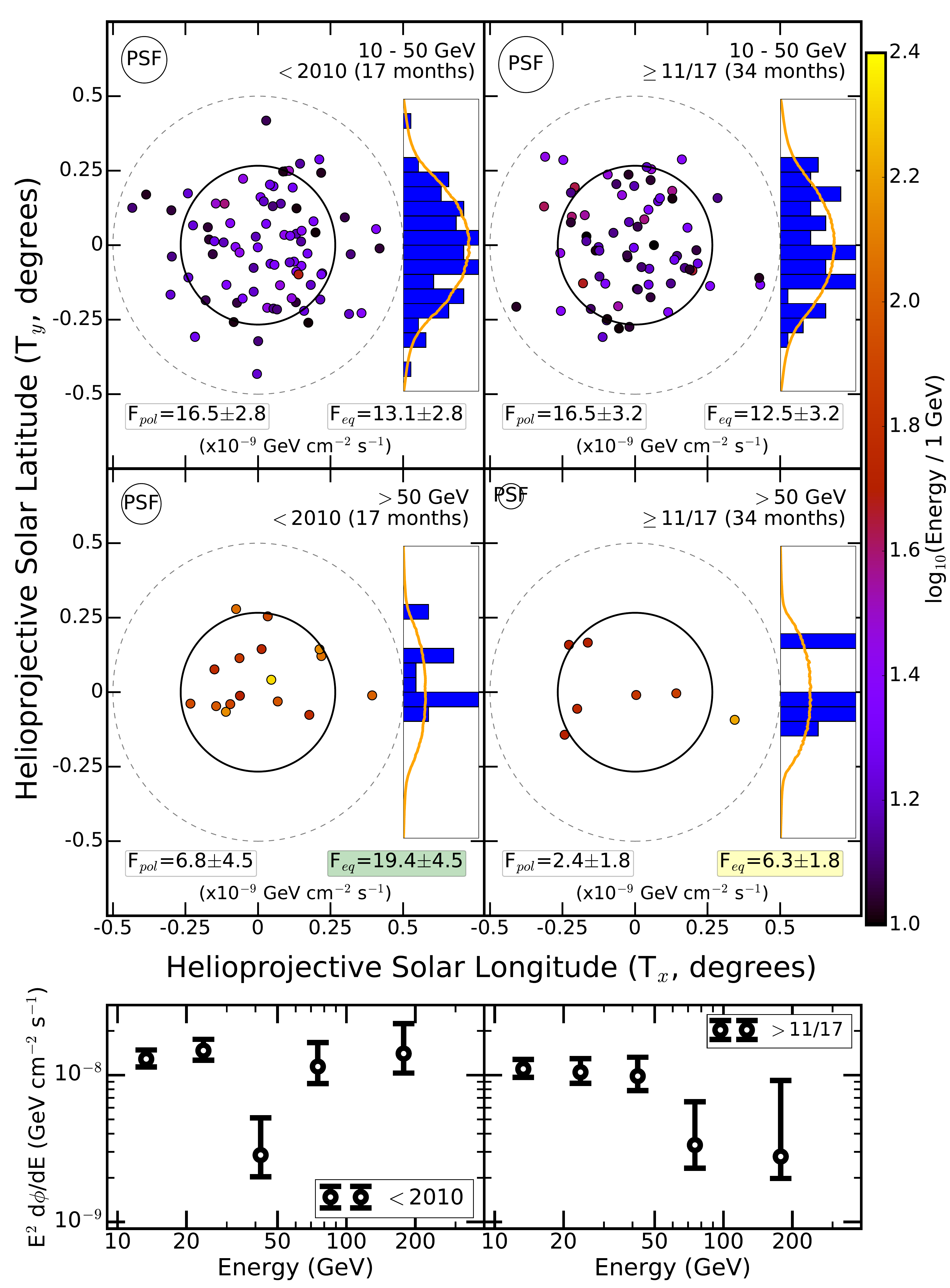}
\caption{{\bf Top:} Locations and energies of high-energy solar-disk $\gamma$-rays from the cycle 23/24 and cycle 24/25 minima in helioprojective coordinates. Data are separated into two temporal bins and two energy bins. The physical solar disk is represented by the solid circle, and the 0.5$^\circ$ ROI by the dashed circle. The 68\% containment region of the PSF for the average $\gamma$-ray is shown in the top left of each panel. The T$_y$ positions of $\gamma$-rays are shown in the histogram, which is compared to the profile expected from isotropic emission smeared by the PSF (orange line). In each panel, we report the flux from the modeled polar and equatorial components, as described in the text. {\bf Bottom:} The $\gamma$-ray spectra during the cycle 23/24 and cycle 24/25 minima.}
\label{fig:helioprojective}
\end{figure*}

Here we focus on comparing Fermi-LAT data from the 2008--2009 solar minimum with data from the 2017--2020 solar minimum. In advance, there are two possible hypotheses:

\begin{itemize}

\item The equatorial and polar emission components depend on the polarity of the heliospheric magnetic field. In this case we might expect a significant shift in the morphology of solar disk $\gamma$-ray emission between the cycle 23/24 and cycle 24/25 solar minima.

\item The equatorial and polar emission components relate to other magnetic phenomena closer to the solar photosphere. In this case, we may expect the $\gamma$-ray emission morphology of these two minima to be similar.

\end{itemize}

We divide the data into four categories, based on the minima to which the event corresponds and the event energy. For the cycle 23/24 minimum, we analyze $\gamma$-ray data between 4 August 2008 (the beginning of the Fermi mission) and 1 January 2010. For the cycle 24/25 minimum, we analyze $\gamma$-ray data between 1 November 2017 and 29 August 2020.  We cut our events into two energy bands, the first spanning from 10--50~GeV and the second for events above 50~GeV. For this analysis, we do not make any cuts on solar flares, as none are observed above our energy threshold. Additionally, we produce no model for the ICS, Moon, or astrophysical backgrounds, noting (as in Ref.~\cite{Linden:2018exo}), that these backgrounds combine to produce only $\sim$1\% of the total flux in our ROI. We do however, make one cut on the data, eliminating all events observed while the Sun lies within 5$^\circ$ of the galactic plane, a choice that significantly decreases the level of astrophysical contamination but only moderately affects the total exposure.

Despite the fact that observations during the current solar minimum span a duration twice as long as the previous minimum (1032 vs.\ 515 days), observations of the current minimum include nearly the same (99\%) exposure, due to the decreased solar exposure of the Fermi-LAT after March 2018.  The present results for solar cycle 23/24 are almost identical to those of Ref.~\cite{Linden:2018exo}, with small fluctuations due to the adoption of the more-recent P8\_V3 IRFs, which have slightly changed the angular uncertainty of several events.

Figure~\ref{fig:helioprojective} shows the morphology of $\gamma$-ray emission across the solar disk. During both solar minima, the emission at energies of 10--50~GeV is nearly uniform across the solar surface, while the emission above 50~GeV shows a prominent equatorial component and little polar emission.

To further probe the polar and equatorial components of the solar disk flux, we follow the procedure of Ref.~\cite{Linden:2018exo}, and fit the $\gamma$-ray emission utilizing a two-component model that includes homogeneous $\gamma$-ray emission across either: (a) the polar half of the solar disk, or (b) the equatorial half of the solar disk. These portions are divided at a helioprojective latitude, T$_y$, of $\pm$0.108$^\circ$. The emission from each component is smeared by the best-fit PSFs for each individual $\gamma$-ray, utilizing the King function model that describes the angular uncertainty of photons in the P8\_V3 source PSF.  

Utilizing this model, we find the best-fit fluxes of the polar and equatorial components, as well as their relative uncertainties.  The polar and equatorial flux uncertainties depict the fractional contributions of each component to the recorded number of photons, and \emph{do not} include the statistical uncertainty of the total $\gamma$-ray flux (i.e., those coming from Poisson fluctuations in the number of photons that were observed). 

Our analysis shows that the $\gamma$-ray morphology during the current and previous solar minima are similar. In both cases, the polar and equatorial $\gamma$-ray fluxes are similar between 10--50~GeV, while equatorial emission dominates at higher energies. For comparison, Ref.~\cite{Linden:2018exo} found that the polar emission was dominant during the remaining solar cycle (2010--2017), where (75.6 $\pm$ 4.1)\% of the emission between 10--50~GeV, and (93.1 $\pm$ 6.8)\% of the emission above 50~GeV, came from polar regions of the solar disk.

Figure~\ref{fig:helioprojective} also shows the total $\gamma$-ray flux and uncertainty in four logarithmic energy bins over 10--100 GeV, along with one additional bin that includes the data between 100--316. GeV (encompassing all high-energy events in this analysis). For each bin, we include uncertainties based on the Poisson fluctuations in the $\gamma$-ray count in each bin (producing the asymmetric error bars that are observed). 

The $\gamma$-ray spectrum during the current solar minimum is somewhat softer than the previous minimum. We find only 7 $\gamma$-rays with energies above 50~GeV during the current minimum, compared to 16 events during the previous minimum, despite almost identical exposures. Moreover, events observed during the current minimum have lower energies (73.4 $\pm$ 37.3 GeV), compared to the previous minimum (97.4 $\pm$ 39.8~GeV). This combination leads to a factor of $\sim$3 reduction in the $\gamma$-ray energy flux from the last minimum to the present one.  Indeed, the $\gamma$-ray flux during the cycle 24/25 minimum barely exceeds the solar maximum flux, which had an energy flux of \mbox{6.9 $\times$10$^{-6}$~MeV~cm$^{-2}$ s$^{-1}$}, compared to the solar minimum value of \mbox{8.7 $\times$10$^{-6}$~MeV~cm$^{-2}$ s$^{-1}$}. Upcoming high-energy observations by the HAWC telescope, with a much larger effective area in the high-energy regime, will be important for understanding the high-energy disk emission during solar minimum. Observations during the solar maximum found no evidence for significant $\gamma$-ray emission above an energy threshold of $\sim$1~TeV~\cite{Albert:2018vcq}.

Finally, we note that the $\sim$30-50~GeV spectral dip (first identified in~Ref.~\cite{Tang:2018wqp} and clearly detected during the first solar minimum here) is not observed during the most recent minimum. In fact, the 30-50~GeV flux from the current minimum significantly exceeds the flux from 2008--2010. The reason is unknown. One possibility is that the spectral dip depends on polarity, and disappears during positive polarities. Another possibility is that the energy-range of the spectral dip shifts. Notably, if the dip moved to higher energies (e.g., 50--80~GeV) it would also explain the different high-energy fluxes. At present, there is no model that favors either scenario.


\section{Discussion and Conclusions}
\label{sec:conclusions}

The solar disk is one of the brightest $\gamma$-ray sources observed by the Fermi-LAT.  Although the basic cause of the $\gamma$-ray emission is known to be bombardment by hadronic cosmic rays, many basic facts are unexplained.  Compared to model predictions, the observed $\gamma$-ray emission is surprising: its luminosity is higher, its spectrum is harder (and features a mysterious dip), its time variation is stronger, and its morphology of the surface emission is less uniform~\cite{Ng:2015gya, Tang:2018wqp, Linden:2018exo}.   A key way forward is to seek new observational clues.  In other work, we have sought to extend the spectrum to higher energies through searches with HAWC~\cite{Nisa:2019mpb, Albert:2018vcq}.  Here we include a push to lower energies in the observational analysis.  In a separate paper~\cite{Zhu2020_TBS}, we will provide new theoretical work at lower energies.  Understanding the Sun is important in its own right, and is also an input to other physics, for example, tests of dark matter~\cite{1985ApJ...296..679P, Peter:2009mk, 2012PhRvD..85l9905S, Ng:2017aur, Leane:2017vag, Edsjo:2017kjk, Albert:2018jwh, Niblaeus:2019gjk,  Cuoco:2019mlb, Mazziotta:2020foa}.

In this paper, we have produced detailed observations of the solar $\gamma$-ray flux over a full solar cycle, spanning an energy range between 100~MeV and 100~GeV, with the first measurements of data below 1 GeV since 2011~\cite{2011ApJ...734..116A}.  Utilizing a novel methodology to measure and subtract astrophysical backgrounds, we have made new high-precision measurements of the solar-disk  $\gamma$-ray spectrum, temporal variation, and morphology. The solar disk $\gamma$-ray spectrum is significantly harder ($dN/dE \sim E^{-2.0}$ at low eneriges) compared to the cosmic-ray spectrum  ($dN/dE \sim E ^{-2.7}$).

Our observations indicate that the dominant temporal trend is a strong anticorrelation of solar $\gamma$-ray emission with solar activity (measured e.g., via sunspot number). The $\gamma$-ray flux increases by approximately a factor of 2 during solar minimum and falls smoothly as the Sun approaches solar maximum. Notably the fall and rise of the solar disk $\gamma$-ray flux is symmetric around the solar maximum, with no evidence of a deviation based on the polarity of the heliospheric magnetic field. This trend matches previous observational data~\cite{Ng:2015gya}, but runs contrary to modeling efforts aimed at understanding solar $\gamma$-ray emission~\cite{Li:2020gch}. 

Further, we observe that the 11-yr variation has no dependence on energy.  And, in the low-energy spectra (100 MeV to 10 GeV), we see no significant differences other than the overall variation.  These are clues that the controlling reason for the variability is magnetic fields in the solar atmosphere as opposed to cosmic-ray modulation (contrary to previous models~\cite{Omodei:2018uni}).  The is supported by hints of features in the energy spectra and time profiles shown above, as the complexity of those fields would introduce a variety of energy and time scales.

Interestingly, our analysis finds no additional evidence for time variability beyond the 11-yr solar cycle (excluding contributions from specific solar flares). Three separate analyses have been performed to test this phenomenon: (1) yearly analyses where the full morphology of ICS emission was allowed to float, (2) near-monthly analyses where the morphology of ICS emission was fixed to the global value, (3) a novel auto-correlation function approach that searched for general time variabilty.  

In addition to our focus on low energies, we also conduct a new study of the highest-energy Fermi-LAT data, comparing the $\gamma$-ray emission observed during the 2007--2010 solar minimum and the 2017--2020 solar minimum. Because these solar minima both have low-solar activity but have opposite magnetic polarities, comparing the $\gamma$-ray signal during these periods can allow us to constrain physical models for high-energy solar $\gamma$-ray emission. We obtain several important results: (1) The flux and spectrum below $\sim$30~GeV are compatible between the two minima, indicating that the Sun's polarity does not significantly affect the $\gamma$-ray signal through most of the Fermi-LAT energy range, (2) restricting ourselves to the best localized $\gamma$-rays above 10~GeV, we find a similar preference for equatorial, rather than polar, emission during each minimum, indicating that the magnetic polarity also does not affect the location at which cosmic-rays preferentially interact with the solar surface, (3) we find no evidence for the 30--50~GeV ``dip" during the current solar minimum, a feature that was most pronounced during the 2008 solar minimum~\cite{Tang:2018wqp}, and (4) we find less evidence for emission above 100~GeV during the current solar minimum (an observation of 1 $\gamma$-ray, compared to 8 $\gamma$-rays in the previous minimum, with similar exposures). However, we note that the statistics are not sufficient to rule out the possibility that these results are consistent.

Finally, we note that the large, time-dependent dataset produced by our novel methodology allows us to correlate solar-disk $\gamma$-ray emission against nearly any time-dependent solar observable. For example, studies correlating solar $\gamma$-ray emission with the detection of coronal mass ejections, helmet streamers, coronal holes, and other solar phenomenon are ongoing. Furthermore, we can reverse the process described in this paper to remove disk emission and measure the spectrum, time variability, and morphology of the ICS halo surrounding the solar position, for which results will be released in a forthcoming publication~\cite{Linden2021_TBS}.


\section*{Acknowledgements}

We thank Gulli Jóhannesson, Igor Moskalenko, Mehr Un Nisa, Elena Orlando, and Andy Strong for helpful comments. TL is partially supported by the Swedish Research Council under contract 2019-05135, the Swedish National Space Agency under contract 117/19 and the European Research Council under grant 742104.  JFB, BZ and GZ were supported by NSF grant Nos.\ PHY-1714479 and PHY-2012955. AGHP was supported by NASA Grant Nos.\ 80NSSC18K1728 and 80NSSC19K1519. BB was supported by NASA Grant No. 80NSSC19K1709. TL, BZ, and GZ were additionally supported by NASA Grant Nos.\ 80NSSC18K1728 and 80NSSC19K1519.  


\begin{figure*}[t!]
\includegraphics[width=0.98\columnwidth]{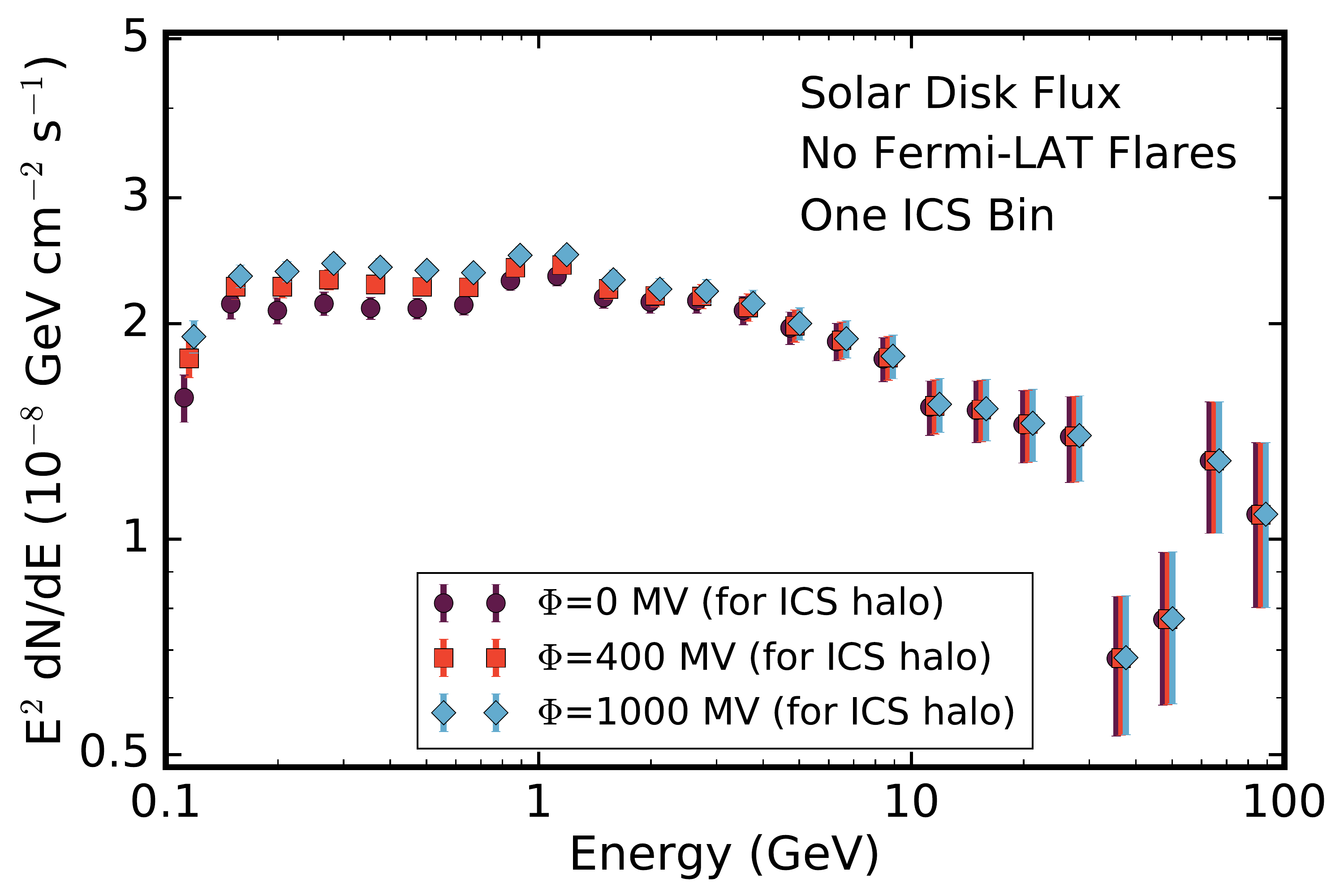}
\includegraphics[width=0.98\columnwidth]{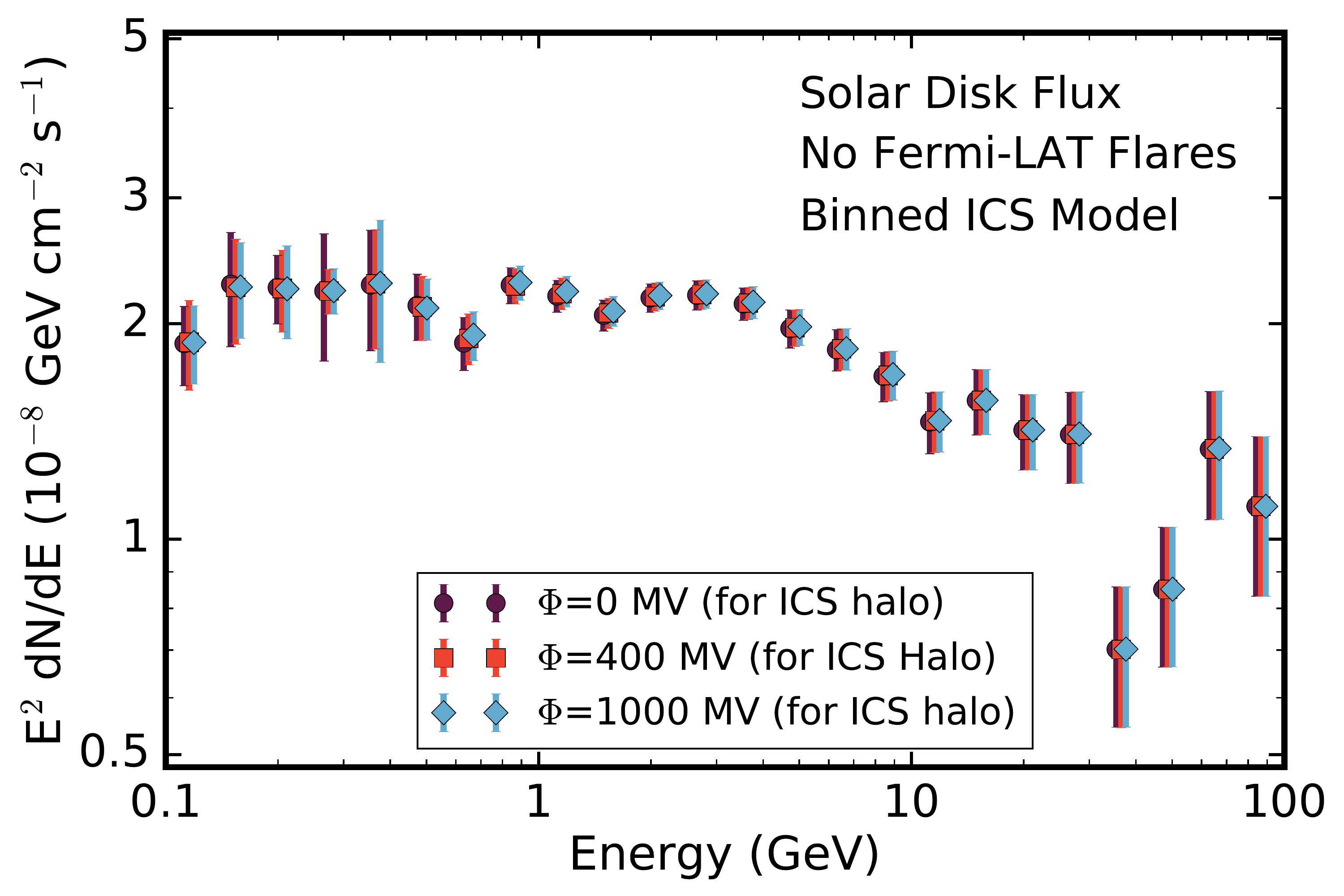}
\caption{{\bf (Left):} Solar-disk $\gamma$-ray flux calculated using three different values for the solar modulation potential of the ICS template. Because ICS emission surrounds the Sun, is brighter than the disk (but with a lower surface brightness), and has a mean $\gamma$-ray distance several degrees from the Sun, different modulation models affect the observed disk flux. The effect is most significant at low energies, where the Fermi-LAT angular resolution is poor. {\bf (Right):} Same, but dividing the ICS model into seven angular bins spanning \{0--1, 1--2, 2--4, 4--6, 6--8, 8--10, 10--45\}$^\circ$, with the normalization of each bin allowed to float independently. Effects from our choice of solar modulation model are removed, and much greater latitude is given to the ICS morphology, producing conservative uncertainty bars that encompass both statistical and systematic errors.}
\label{fig:icsnorm}
\end{figure*}

\newpage


\appendix

\section{Effect of Inverse-Compton Models}
\label{app:ics}

\begin{figure*}[t]
\includegraphics[width=0.98\columnwidth]{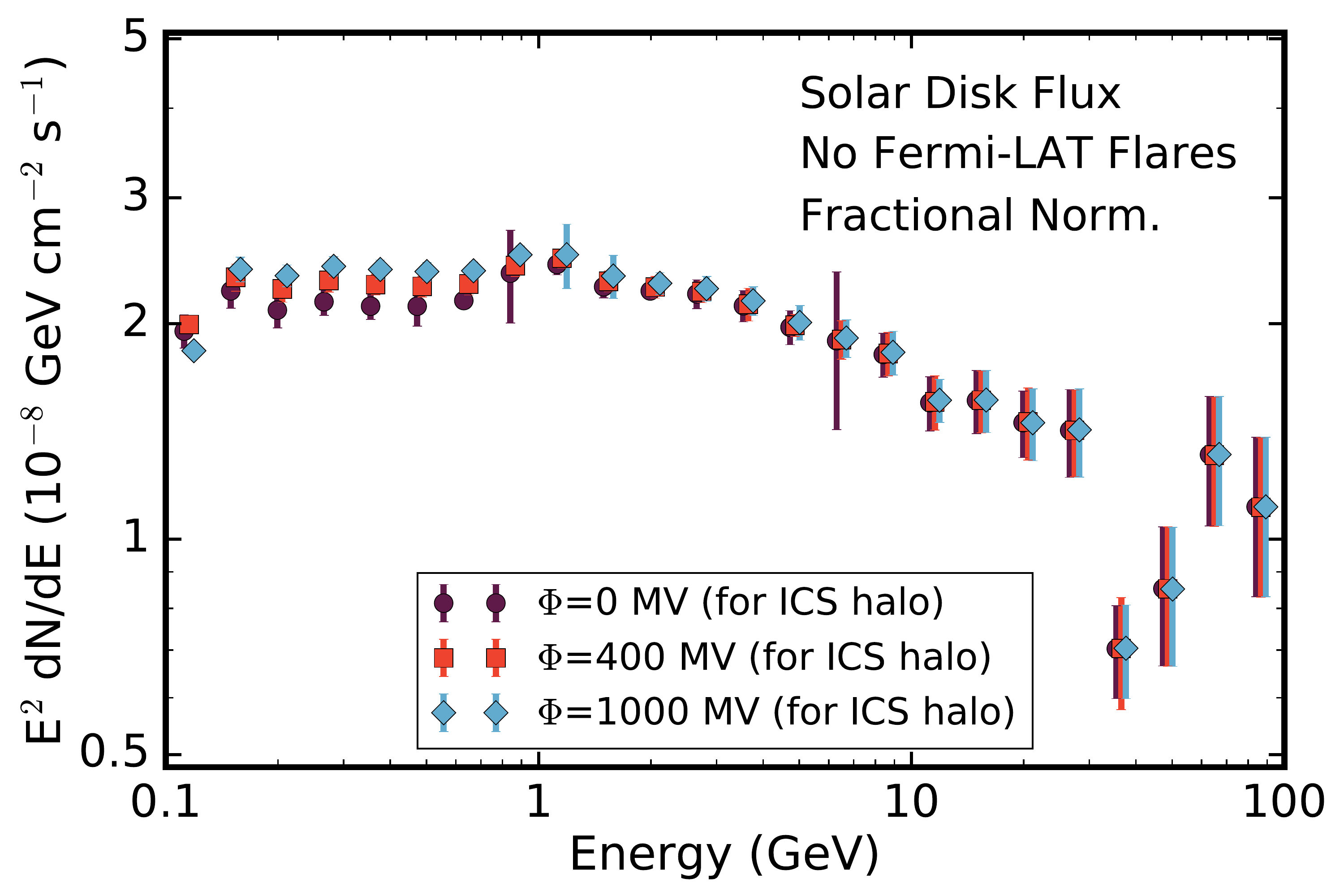}
\includegraphics[width=0.98\columnwidth]{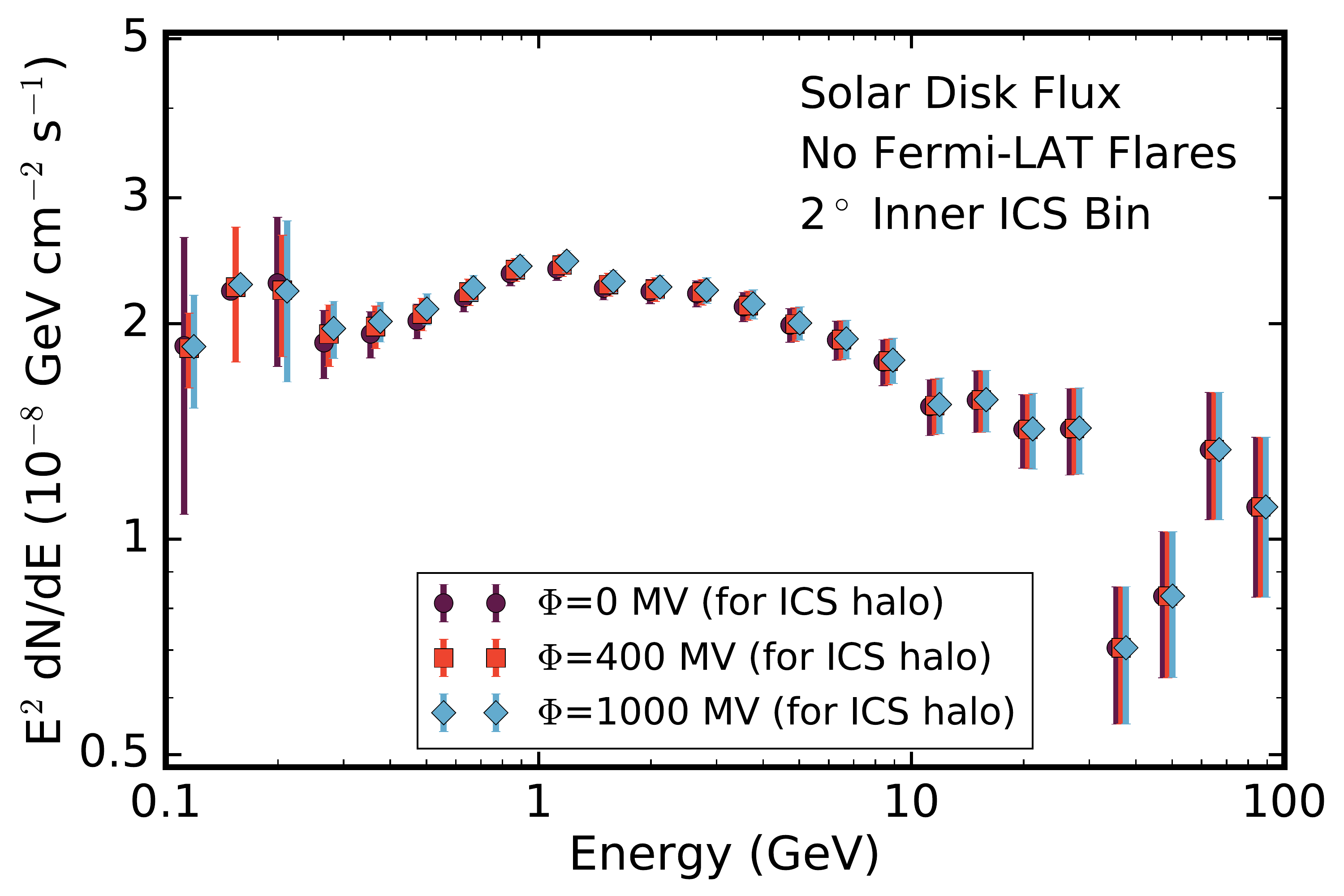}
\caption{{\bf (Left):} Same as Figure~\ref{fig:icsnorm} (right), except that the intensity normalization of each ICS angular bin is forced to be smaller than the larger angular bin surrounding it. This mimics the physical intuition that solar modulation can only prevent cosmic rays from reaching the Sun. The best-fit disk flux in this model is consistent with our default model, but the error bars are much smaller, and are not statistically consistent for different values of $\phi$. {\bf (Right:)} Same as Figure~\ref{fig:icsnorm} (right), except the two inner bins are replaced by a single bin that spans 0--2$^\circ$. This removes some degeneracy between the disk and ICS flux between 0.3--1~GeV, at the cost of inducing a slight spectral feature at the lowest energies.}
\label{fig:icsnorm_2}
\end{figure*}

In addition to emission from the disk, the Sun produces a bright $\gamma$-ray flux via the interaction of cosmic-ray electrons with solar photons. The total flux of this component is a factor of a few brighter than the disk. However, the surface brightness of the ICS halo is much lower, with significant emission detected out to more than 20$^\circ$ from the Sun. Moreover, kinematic constraints (the similar directions of outgoing solar photons and cosmic-ray electrons) significantly inhibit ICS very close to the Sun. In this Appendix, we utilize these facts to differentiate between disk and halo emission. 

If the Sun had no magnetic fields, the ICS halo morphology and energy spectrum would be trivial to calculate. Cosmic-ray electrons are produced by interstellar processes, and are homogeneous in the solar system. The energy density of solar photons falls as 1/r$^2$. However, solar winds and magnetic fields affect the propagation of electrons throughout the heliosphere. The strength of this interaction depends on the microphysics of heliospheric  plasma~\cite{Vittino:2017fuh, Boschini:2019ubh}. However, the bulk effect of solar modulation can be modeled based on the phase of the solar cycle, the tilt angle of the heliospheric current sheet, and the polarity of the heliospheric magnetic field~\cite{Cholis:2015gna, Cholis:2020tpi}. Many models simplify the physics further, modeling solar modulation using a charge-, time- and rigidity- dependent potential that cosmic-rays must climb before encountering their target~\cite{Kuhlen:2019hqb}. The value of this potential is fit to experiments.

Solar modulation also affects the spectrum and flux of the solar disk by modulating the cosmic-ray protons that impingge on the Sun. We discuss this in detail in~\cite{Zhu2020_TBS}. However, the morphological impact of this modulation is small. At the level of the Fermi-LAT PSF, we can model the Sun as a homogeneous disk. The morphology of ICS emission, however, depends sensitively on the choice of modulation parameters. 

Uncertainties in the ICS make it difficult to model the solar disk. ICS emission is brighter than the disk, making its accurate modeling important. Because the majority of ICS $\gamma$-rays are located several degrees from the Sun, the likelihood model fits the ICS intensity (flux per solid angle) far from the Sun, and then extrapolates that result close to the Sun based on the assumed ICS morphology. Thus, ICS models with different modulation potentials will produce different solar disk fluxes. The effect is most pronounced at low energies, where the poor PSF of the Fermi-LAT smears out the small-scale differences between ICS and disk morphologies. In this paper, we concern ourselves only with the impact of ICS mismodeling on the calculated solar disk flux and spectrum. We leave our analysis of the ICS template itself to future work~\cite{Linden2021_TBS}.

Figure~\ref{fig:icsnorm} shows the solar disk spectrum for various ICS emission models. First (left), we plot the solar disk emission assuming ICS morphologies produced using a single potential over its full extent (0--45$^\circ$). We note two important results. First, above an energy of $\sim$1~GeV, the degeneracy between our ICS model and the solar disk flux disappears entirely. In this range, the Fermi-LAT PSF is sufficient to differentiate between the disk and ICS halo. Second, we note a significant degeneracy at lower-energies, producing a factor of $\sim$10-20\% uncertainty in the solar disk flux in the lowest energy bands.

\begin{table}
\begin{tabular}{|c|c|c|c|}
  \hline
Energy (GeV) & $\Delta$ LG($\mathcal{L}$) $\phi_0$ & $\Delta$LG($\mathcal{L}$) $\phi_{400}$ & $\Delta$LG($\mathcal{L}$) $\phi_{1000}$ \\
\hline
  0.10 -- 0.13 & 1.35 & 0.94  & \cellcolor{green!25}0.0 \\
  0.13 -- 0.18 & \cellcolor{green!25} 0.0 & 0.74 & 1.46 \\
  0.18 -- 0.24 & 5.84 & 3.00 & \cellcolor{green!25} 0.0 \\
  0.24 -- 0.32 & 5.96 & 3.55 & \cellcolor{green!25} 0.0 \\
  0.32 -- 0.42 & 7.55 & 5.01 & \cellcolor{green!25} 0.0 \\
  0.42 -- 0.56 & 0.61 & 2.23 & \cellcolor{green!25} 0.0 \\
  0.56 -- 0.75 & \cellcolor{green!25} 0.0 & 5.70 & 7.71 \\
  0.75 -- 1.00 & \cellcolor{green!25} 0.0 & 3.79 & 4.17 \\
  1.00 -- 1.33 & \cellcolor{green!25} 0.0 & 4.67 & 6.68 \\
  1.33 -- 1.78 & \cellcolor{green!25} 0.0 & 3.67 & 5.35 \\
  1.78 -- 2.37 & \cellcolor{green!25} 0.0 & 3.67 & 6.43 \\
  2.37 -- 3.16 & \cellcolor{green!25} 0.0 & 1.31 & 1.96 \\
  3.16 -- 4.22 & \cellcolor{green!25} 0.0 & 0.90 & 1.66 \\
  4.22 -- 5.62 & \cellcolor{green!25} 0.0 & 0.21 & 0.39 \\
  5.62 -- 7.50 & \cellcolor{green!25} 0.0 & 0.06 & 0.12 \\
  7.50 -- 10.0 & 0.17 & 0.10 & \cellcolor{green!25} 0.0 \\
  10.0 -- 13.3 & \cellcolor{green!25} 0.0 & 0.08 & 0.19 \\
  13.3 -- 17.8 & \cellcolor{green!25} 0.0 & 0.16 & 0.38 \\
  17.8 -- 23.7 & \cellcolor{green!25} 0.0 & 0.04 & 0.09 \\
  23.7 -- 31.6 & \cellcolor{green!25} 0.0 & 0.12 & 0.29 \\
  31.6 -- 42.2 & \cellcolor{green!25} 0.0 & 0.10 & 0.24 \\
  42.2 -- 56.2 & \cellcolor{green!25} 0.0 & 0.08 & 0.21 \\
  56.2 -- 75.0 & \cellcolor{green!25} 0.0 & 0.07 & 0.18 \\
  75.0 -- 100. & \cellcolor{green!25} 0.0 & 0.14 & 0.34 \\
\hline
\end{tabular}
       \caption{Change in log-likelihood fit for three choices of the solar modulation potential ($\phi_{0}$~=~0~MV, $\phi_{400}$~=~400~MV, $\phi_{1000}$~=~1000~MV). The $\Delta$LG($\mathcal{L}$) is compared to the best fit model, thus one column is always 0 (shaded green) by construction. We find that strong solar modulation is preferred for $\gamma$-ray energies below $\sim$500~MeV, but the model quickly shifts to prefer very small modulation potentials at higher energies. }
    \label{tab:fits}
\end{table}

The direction of this degeneracy is intuitive. The average ICS $\gamma$-ray is located 10--15$^\circ$ away from the Sun --- and the critical region for fitting the ICS template (in terms of the log-likelihood) is 5--10$^\circ$ from the Sun. The ICS model is driven by fits within these regions. Because models with small modulation potentials have a large fraction of their flux close to the disk, the fit will produce a bright ICS flux near the Sun. Models with large modulation potentials have a small flux close to the disk, and produce a dim ICS flux near the Sun. The solar disk template is the only parameter capable of responding to this change, and its flux floats accordingly. As the energy increases, the degeneracy disappears, because all ICS models fall off very close to the Sun due to kinematic considerations (when the angle between the momentum of a cosmic-ray electron and a solar photon is 0, the ICS cross-section is 0).

\begin{figure}[t]
\includegraphics[width=0.98\columnwidth]{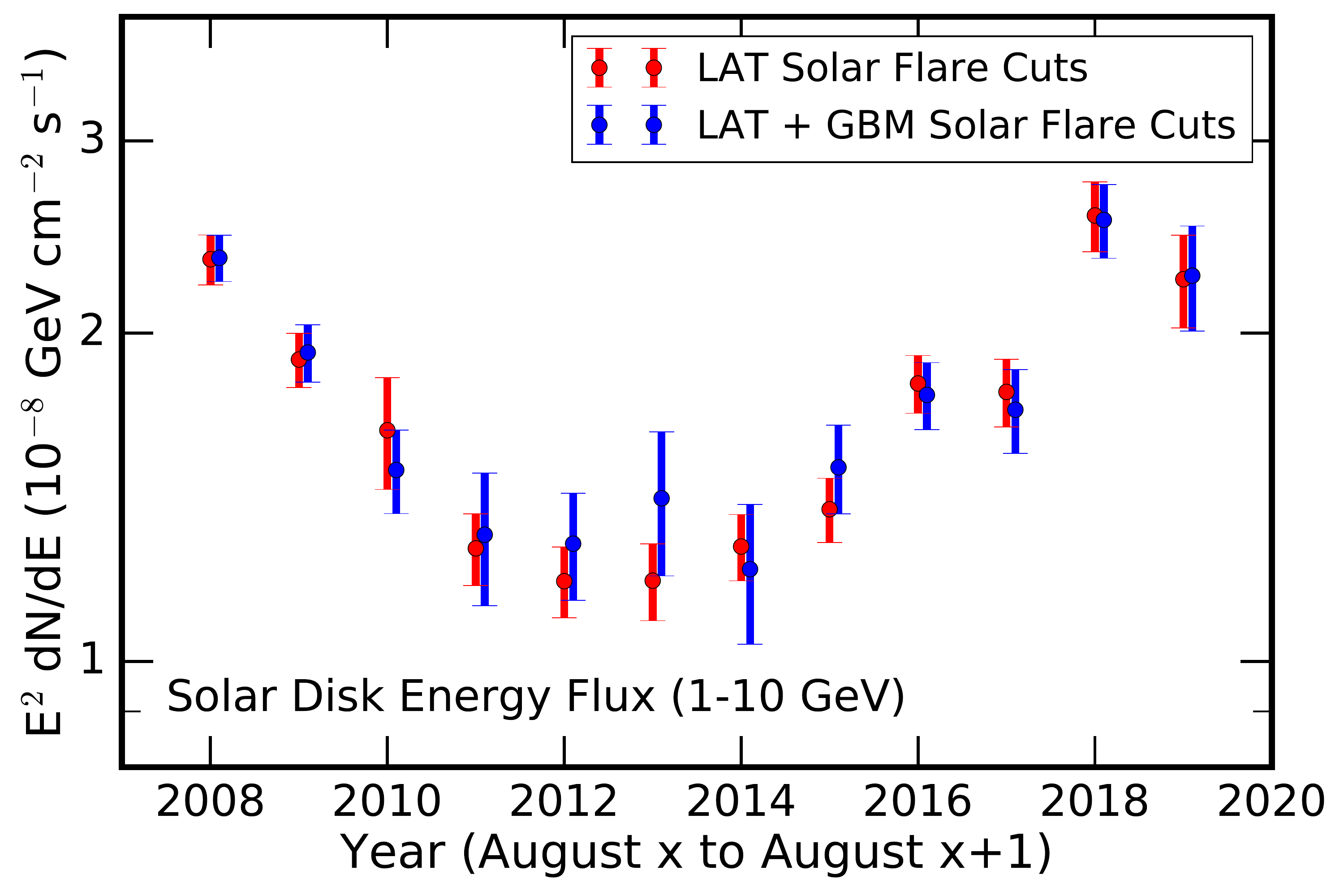}
\caption{Same as Fig.~\ref{fig:time_variability}, but showing results for both our standard cut on Fermi-LAT detected flares, and our conservative cut on both LAT- and GBM- selected flares. In each year, the $\gamma$-ray fluxes from both cuts are similar, verifying that sub-threshold flares are not correlated (or anti-correlated) with $\gamma$-ray emission above 1~GeV.}
\label{fig:time_variability_GBM}
\end{figure}

There are many ways to account for this degeneracy. In Figure~\ref{fig:icsnorm} (right), we show our default choice. We divide the ICS template into 7 angular bins, spanning \{0--1, 1--2, 2--4, 4--6, 6--8, 8--10, 10--45\}$^\circ$. The ICS template normalizations are allowed to float independently in each bin (but are fixed to be non-negative), allowing the ICS flux near the Sun to float independently of more distant regions. We note three results: (1) All ICS models produce a consistent measurement of the disk flux, implying that the ICS morphology \emph{within} each bin does not affect the result. (2) The flux and uncertainty in the disk emission above $\sim$1~GeV is unchanged, indicating that our modeling does not affect higher energy emission. (3) The error bars on the low-energy disk flux have increased, indicating that there is a significant degeneracy between the disk flux and the normalization of the inner-most ICS ring.

In Figure~\ref{fig:icsnorm_2}, we test two alternative methods. First (left), we utilize the same binning, but force the normalization factor of each inner ICS bin to be smaller than that of the larger ICS bin surrounding it. This is theoretically motivated because solar modulation primarily repels electrons --- with stronger effects closer to the Sun. In this case, we find much smaller statistical uncertainties for each choice of modulation potential. This indicates that the largest uncertainty in the disk flux comes from scenarios where the ICS model adopts a ``saw-tooth" configuration in the angular bins. However, in this method, the low-energy fits for different modulation models are not consistent, indicating that there is a residual systematic error in our model of the disk flux. For this reason, our default model allows the normalization of each ICS bin to float independently --- as we believe that the statistical errors produced in that fit more accurately represent the true uncertainty.

In Figure~\ref{fig:icsnorm_2} (right), we again repeat our default analysis (allowing the flux in each bin to float to any non-negative value). However we combine the two inner bins to form a single bin spanning 0--2$^\circ$. This decreases the degeneracy of the ICS template at moderate energies (between 0.3--1~GeV), decreasing the uncertainty in the disk flux at those energies. The cost of this choice is the appearance (at low-significance) of a bump at $\sim$0.25~GeV, which is produced when the angular resolution of the template becomes too poor and the disk emission becomes degenerate with the larger ICS bin.

Overall, we find that these methods all provide rigorous fits to the solar disk flux. Models with a single ICS template provide results with excellent statistical precision, but include systematic errors that depend on the model choice. However, our default binning scheme produces results that: (i) do not depend on the choice of the modulation potential (ii) have systematic uncertainties that are small compared to their conservative (and quantified) statistical uncertainties. Thus, we believe that our default analysis accurately represents our state of knowledge regarding the low-energy solar disk flux, though we note that the error bars may be conservative.

We note that this method differs from traditional Fermi-LAT analyses, which focus on testing multiple physical models for the ICS emission, and then minimizing the log-likelihood fit for each model. In this analysis, however, such a technique is sub-optimal, because the the log-likelihood can be highly influenced from the fit of the ICS template far from the solar disk. We reserve a full likelihood analysis of the ICS morphology to a later paper~\cite{Linden2021_TBS}.

However, we note that our analysis provides significant evidence for a modulation potential with a non-trivial rigidity dependence. In Table~\ref{tab:fits}, we show the $\Delta$\rm{LG}($\mathcal{L}$) in 24 energy bins between 0.1--100~GeV for models with a single ICS component. At low $\gamma$-ray energies ($\lesssim$0.5~GeV), our fits prefer very high modulation potentials, which remove any low-energy electrons from regions near the Sun. However, at higher energies, this situation reverses and our models prefer modulation potentials near 0. The rapid shift (occurring over a single energy bin), may signal either an abrupt change in the effect of the heliospheric magnetic field on cosmic-rays, or may alternatively indicate a systematic error in the differentiating the solar disk and ICS templates at low energies. Even if the abruptness of the change is unexpected, our model provides strong evidence for a important solar modulation effect at low energies, and a very small effect in the higher energies. We will provide more details concerning this scenario in Ref.~\cite{Linden2021_TBS}.


\begin{figure*}[t]
\includegraphics[width=1.98\columnwidth]{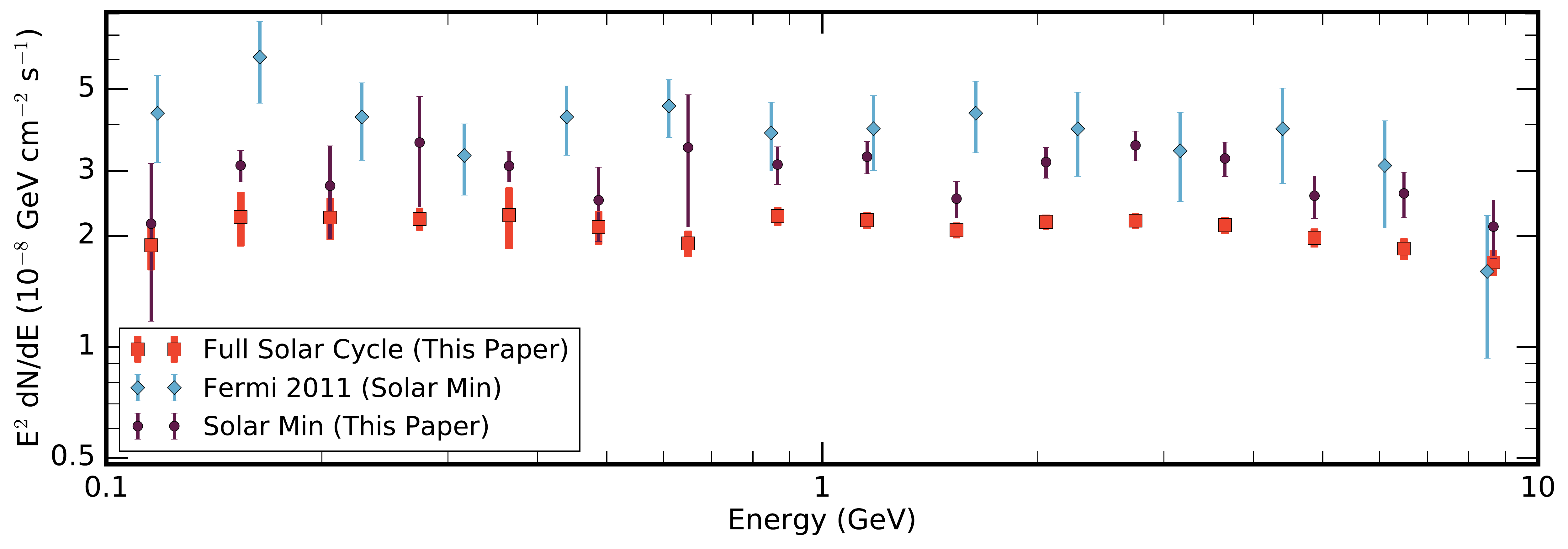}
\caption{Comparison of the solar-disk $\gamma$-ray flux determined in our analysis (orange), with the most recent results on the solar-disk $\gamma$-ray flux below 1~GeV~(blue, Fermi2011). Also shown are the best-fit $\gamma$-ray fluxes utilizing our analysis technique over the solar minimum time period which closely matches the analysis period observed in~Fermi2011 (brown). Datasets from this paper are binned at 8 logarithmic energy bins per decade, while results from Fermi2011 are binned at 7 logarithmic energy bins per decade.} 
\label{fig:fermi2011}
\end{figure*}

\section{Accounting for Bias in Fermi-LAT/GBM Cut}
\label{app:flares}

In Sec.~\ref{subsec:flux}, we utilized two different cuts on solar flare activity intended to isolate the steady state $\gamma$-ray emission from the solar disk. The first cut removed data from time-slices corresponding to events in the Fermi-LAT significant flares list. This removed only 0.3\% of the total solar exposure, but removed the vast majority of flare activity. We additionally attempted a more liberal cut that removed not only the Fermi-LAT significant flares, but also all low-energy flares detected by the Fermi-GBM. This cut removed nearly 40\% of the total data, and strongly limited the potential contribution from sub-threshold solar flares to the solar $\gamma$-ray flux.

In Figure~\ref{fig:spectrum}, we showed these methodologies produced similar $\gamma$-ray spectra, indicating that removing just a handful of the brightest solar flares was sufficient to clean the dataset of transient flare event. This conclusion was further strengthened by the lack of small-scale time-variability features which would be indicative of transient flare activity.

One confusing result from Figure~\ref{fig:spectrum} is the relative enhancement of the $\gamma$-ray flux above 1~GeV during periods when both LAT/GBM flares are removed. In the main text we noted that this is a correlation, but not a causation -- GBM flares predominantly occur during periods of high-solar activity, and periods with high solar activity have small steady-state $\gamma$-ray fluxes. That is, our choice to remove GBM cuts biased our solar exposure towards time periods where the solar disk happened to be bright.

To verify this, in Figure~\ref{fig:time_variability_GBM} we show the $\gamma$-ray flux above 1~GeV using both our LAT and LAT/GBM time cuts. We further divide the data into yearly components, following the methodology from the main text. We find that in each period, the LAT and LAT-GBM cuts produce similar $\gamma$-ray fluxes. This provides further evidence supporting our assertion that the removal of GBM flares does not affect the steady state $\gamma$-ray spectrum.


\section{Comparison with Fermi-LAT Collaboration \cite{2011ApJ...734..116A}}
\label{app:fermi2011}

The most recent previous analysis of the solar disk $\gamma$-ray flux below 1~GeV was performed by the Fermi-LAT collaboration in 2011~\cite{2011ApJ...734..116A} (hereafter, Fermi2011). Here, we compare our results and discuss our interpretation of their differences. 

In Figure~\ref{fig:fermi2011}, we show our spectrum compared to Fermi2011. Two results immediately stand out. The first is the significantly smaller error bars in our flux determinations, a result of the nearly 100-fold increase in exposure enabled by a combination of a longer observation time, as well as the elimination of cuts based on nearby bright sources, the galactic latitude of the Sun, and the position of the Moon.

The second difference is the factor of $\sim$2 flux offset between both observations. This is primarily due to the time-variability of the disk $\gamma$-ray flux (first observed in~\cite{Ng:2015gya} and described in detail in Section~\ref{subsec:variability} here). In Figure~\ref{fig:fermi2011}, we also show a ``Solar Min" flux computed over the period from 4 August 2008 to 1 January 2010 (similar to the 4 August 2008 to 4 February 2010 cut used in~Fermi2011). This corrects the majority of the offset, in particular above 1~GeV, where our results closely agree with those of Fermi2011. We note the improved statistical precision from our analysis, even in data taken over the same time period.

At low energies, we note a small offset between Fermi2011 and our result, particularly near $\sim$160~MeV, where a small bump appears in Fermi2011 which does not appear in our data. We also note that our solar-minimum analysis occasionally has error bars that are similar to, or slightly exceed that of Fermi2011. This is due to the significant freedom given to the normalization of the inner-ICS profile in our analysis -- which implies that our uncertainties include both statistical and systematic uncertainties. If we were to use an analysis that was restricted to uniform or monotonic ICS profiles (based on a simple modulation potential) we would obtain results with significantly smaller uncertainties (see e.g., Figure~\ref{fig:icsnorm_2}).

\section{High-Energy Time Variability}
\label{app:highenergyvariability}

\begin{figure}[t]
\includegraphics[width=0.98\columnwidth]{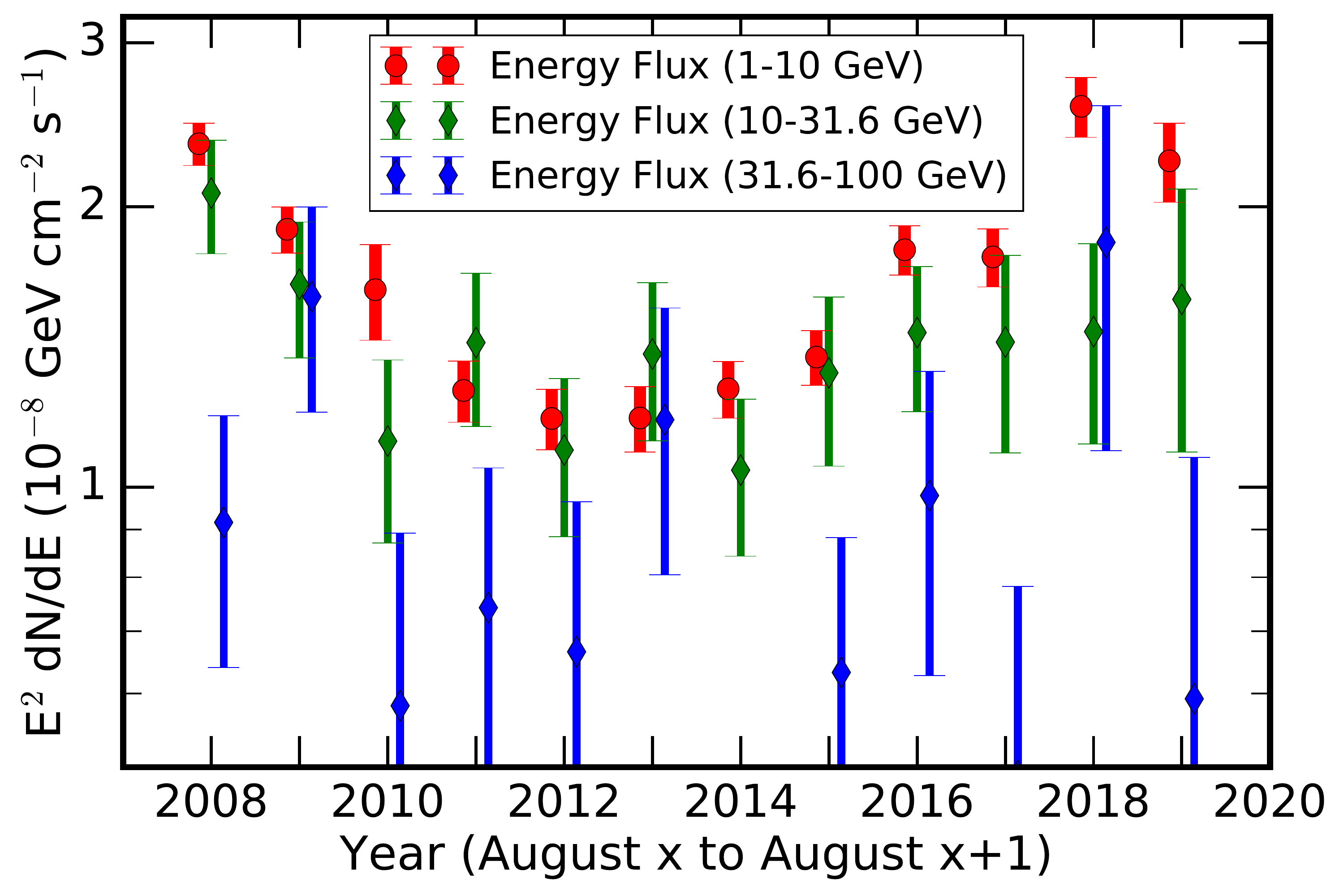}
\caption{Same as Figure~\ref{fig:time_variability}, except utilizing energy bins of 1--10~GeV (red) and 10--31.6~GeV (green) and 31.6--100~GeV (blue). The points are slightly offset for clarity.  We find strong evidence that the 11-yr variability, previously shown to anticorrelate with sunspot number~\cite{Ng:2015gya}, continues up to an energy of $\sim$30~GeV. Note that the highest energy datapoint in 2014 has a best-fit flux of 0 and falls off of the y-axis of the plot. We find little evidence for periodic variability at higher-energies, but note that this the interpretation of this is extremely difficult, as it depends sensitively on our understanding of the spectral dip (see~Ref.~\cite{Tang:2018wqp} and Figure~\ref{fig:spectrum}.) }
\label{fig:time_variability_appendix}
\end{figure}

In Section~\ref{subsec:variability}, we examined the yearly variability in the disk $\gamma$-ray spectrum, focusing on the energy range of 0.1--10~GeV, where the statistical power is highest. Here we examine evidence for variability at higher energies.

We note three critical results from previous studies. First, Ref.~\cite{Ng:2015gya} searched for variability in the energy bin 10--100~GeV, but found no statistically significant evidence for a time-dependent component (though their results were also consistent with the time-dependence observed in the 1--10~GeV data). Secondly, Ref.~\cite{Tang:2018wqp} found statistically significant evidence for a large-amplitude spectral dip between 30-50~GeV, which is most significant during the 2008--2010 solar minimum (data after this period shows some evidence for a dip, but the amplitude is smaller and the data is not statistically significant on its own). Third, the dedicated analysis of the morphology of photons above 10~GeV~\cite{Linden:2018exo}, found 9 $\gamma$-rays with energies above 100~GeV during solar minima, and no such events during the remainder of the solar cycle. 

The combination of these three facts complicates the search for 10--100~GeV variability. We first note that the low statistics in this energy range ($\sim$500~events over 11~years of data), make it difficult to analyze the data (accounting for background modeling and the ICS component) at eight energy bins per decade. The pronounced spectral dip makes it difficult (and inaccurate) to combine energy bins above and below the dip, as the average energy of photons in each time period (and thus the normalized value of E$^2$dN/dE), is not similar. The lack of any events recorded above 100~GeV during the solar maximum (Years 3--8) make it difficult to assign uncertainties (which would be very large in terms of E$^2$dN/dE) in these energy bins. 

Here, we make the following choices, noting that different choices could produce slightly different results. We re-bin our solar disk and background models to four energy bins per decade, and then combine energy bins spanning from 10--31.6~GeV (below the dip) and 31.6--100~GeV (in/above the dip) in post-processing, assuming an average spectrum in the two ranges of E$^{-2}$. We do not include events above 100~GeV, but note that there is very strong evidence for temporal variability in this energy bin~\cite{Linden:2018exo}.

In Figure~\ref{fig:time_variability_appendix}, we show the resulting time variability in these energy bins, comparing our results to data from the 1--10~GeV energy bin shown in the main text. Our analysis shows some evidence for variability in the 10--31.6~GeV dataset, in particular a statistically significant fall in the solar $\gamma$-ray emission between 2008 and 2012. The evidence for an increasing $\gamma$-ray flux during the more recent solar minimum is shakier, partially due to the effect of the diminished solar exposure since 2018. Between energies of 31.6--100~GeV, we find no strong evidence for solar disk variability, but note that in Ref.~\cite{Tang:2018wqp} this is best interpreted as offsetting effects between a brighter disk flux during solar minimum, and a more pronounced solar dip feature during the same time period. While not shown here, we find that our analysis verifies the unique time-variability of $\gamma$-ray emission above 100~GeV (first identified in Ref.~\cite{Linden:2018exo}), with all $\gamma$-ray events observed at this energy stemming from the solar minimum period.


\bibliography{main}

\end{document}